\documentclass[prd,
nofootinbib,
superscriptaddress
]{revtex4}
\usepackage{amsmath,amssymb,bm,float,mathrsfs}
\usepackage{graphicx}
\usepackage{dcolumn}
\usepackage{color}

\usepackage[colorlinks=true
,urlcolor=magenta
,anchorcolor=blue
,citecolor=blue
,filecolor=magenta
,linkcolor=red
,menucolor=blue
,linktocpage=true
,pdfproducer=medialab
,pdfa=true
]{hyperref}

\usepackage{comment}

\newcommand{\be}{\begin{equation}} 
\newcommand{\ee}{\end{equation}}
\newcommand{\bea}{\begin{eqnarray}} 
\newcommand{\eea}{\end{eqnarray}}

\begin{document}

\title{Violation of slow-roll in non-minimal inflation 
}


\author{Tomo Takahashi}
\email[Email: ]{tomot``at"cc.saga-u.ac.jp}
\affiliation{
Department of Physics, Saga University, Saga 840-8502, Japan
}
\author{Tommi Tenkanen}
\email[Email: ]{ttenkan1``at"jhu.edu}
\affiliation{
Department of Physics and Astronomy, Johns Hopkins University, 
Baltimore, Maryland 21218, USA
}

\author{Shuichiro Yokoyama}
\email[Email: ]{shu``at"kmi.nagoya-u.ac.jp}
\affiliation{
Kobayashi Maskawa Institute, Nagoya University, Aichi 464-8602, Japan
}
\affiliation{
Kavli Institute for the Physics and Mathematics of the Universe (WPI), 
Todai Institute for Advanced Study, University of Tokyo, Kashiwa, Chiba 277-8568, Japan
}

\begin{abstract}
We show that a non-minimal coupling to gravity can not only make some inflationary models consistent with cosmological data, similar to the case of Higgs inflation, but can also invoke slow-roll violation to realize a graceful exit from inflation. In particular, this is the case in models where a destabilizing mechanism that ends inflation should be assumed when the model is minimally coupled to gravity. As explicit examples, we consider the power-law and inverse monomial inflation models with a non-minimal coupling to gravity. While these models are excluded in the minimally coupled case, we show that they can become viable again in non-minimally coupled scenarios. In most scenarios we considered, reheating can be naturally realized via gravitational particle production but that this depends on the underlying theory of gravity in a non-trivial way.
\end{abstract}

\preprint{}

\maketitle

\section{Introduction} 
\label{sec:intro}

It is now widely believed that the Universe previously experienced a period of exponentially fast expansion, called cosmic inflation. Although the inflationary Universe has been established as  a paradigm, its 
detailed mechanism or the underlying model responsible for inflation is not yet fully understood, and a lot of theoretical and observational effort has been made during the past decades to elucidate it. From the observational side, recent precise measurements of the cosmic microwave background (CMB) by the Planck satellite have provided tight limits on some inflationary 
observables such as the amplitude and spectral index of the primordial curvature power spectrum and the tensor-to-scalar ratio \cite{Akrami:2018odb}, 
which have provided support for some models but also excluded many (single-field) models of inflation \cite{Martin:2013tda}. 

However, once one extends the framework to scenarios with multiple fields, non-minimal couplings to gravity, and so on, the predictions of some simple models for the spectral index and the tensor-to-scalar ratio can be modified. In multi-field scenarios such as in the curvaton model \cite{Enqvist:2001zp,Lyth:2001nq,Moroi:2001ct} or modulated reheating scenario \cite{Dvali:2003em,Kofman:2003nx}, a so-called spectator field can also contribute to the generation of primordial fluctuations. In such a case, the predictions for the spectral index and the tensor-to-scalar ratio are modified from the usual case and some inflation models become viable again even if the original single-field model is excluded \cite{Langlois:2004nn,Moroi:2005kz,Moroi:2005np,Ichikawa:2008iq,Enqvist:2013paa,Vennin:2015vfa,Haba:2017fbi}. 
Something similar can happen in models with a non-minimal coupling to gravity. 
For example, 
the quartic chaotic inflation with a non-minimal coupling, or the Higgs inflation model, 
which has been excluded by the Planck data as a single-field model, becomes viable again since the predictions for the spectral index and the tensor-to-scalar ratio get modified due to the existence of a non-minimal coupling between the Higgs field and gravity 
\cite{Bezrukov:2007ep} (see also Refs.~\cite{Spokoiny:1984bd,Futamase:1987ua,Salopek:1988qh,Fakir:1990eg,Amendola:1990nn,Kaiser:1994vs,CervantesCota:1995tz,Komatsu:1999mt} for earlier work on the topic,  Refs.~\cite{Rubio:2018ogq,Tenkanen:2020dge} for recent reviews, and Refs.~\cite{Germani:2010gm, Granda:2011zk, Kamada:2012se} for extended work employing a non-minimal derivative coupling).
See also Refs. \cite{Lerner:2011ge,Bezrukov:2011gp,Kaiser:2013sna,Kallosh:2013tua,Gong:2015qha,Takahashi:2018brt} for predictions of inflationary observables in variants of non-minimally coupled models.

Another issue in constructing models of inflation is the so-called graceful exit problem. In some inflation models, such as the power-law inflation \cite{Abbott:1984fp} and the inverse monomial inflation \cite{Ratra:1987rm} models,
the end of inflation cannot be invoked in the usual way -- by violation of slow-roll -- but one needs some destabilizing mechanism, such as tachyonic instability in the inflaton potential, to end inflation. Although such a mechanism does not necessarily affect the models' predictions for observables, one needs to take care of it for a successful and self-consistent inflationary scenario.

In this paper, we show that even a small non-minimal coupling to gravity can also help to end inflation even if one considers models such as the power-law inflation and the inverse monomial inflation models in which the end of inflation cannot be realized by slow-roll violation when the inflaton is minimally coupled to gravity. While this conclusion is not particularly surprising
and has been suggested in the literature (see, {\it e.g.}, Ref.~\cite{Tashiro:2003qp}),
 at the same time -- and more importantly -- the {\it same} non-minimal coupling can also make the spectral index $n_s$ and the tensor-to-scalar ratio $r$ consistent with cosmological data such as those obtained by Planck in spite of the fact that the original models are excluded by the current data as minimally coupled single-field models. 
As we will show, this is a non-trivial requirement, and it
facilitates model building of the inflationary Universe, especially in the case of models with an extended gravity sector.

Another important issue in inflationary cosmology is reheating. Even if a graceful exit from inflation is realized, the Universe still has to be reheated to become radiation dominated by the time of big bang nucleosynthesis (BBN). 
We argue that while the usual mechanism for reheating where the inflaton field oscillates about the minimum of its potential and decays into particles cannot be realized in the models we consider, in most of our scenarios reheating can be realized via gravitational particle production~\cite{Ford,Starobinsky:1994bd}.
This is made possible due to a kination epoch which generically follows the inflationary period in the models we consider. During a kination epoch, the energy density of the inflaton field $\phi$ scales as $\rho_\phi \propto a^{-6}$, with $a$ being the scale factor, and therefore, it decays faster than that of radiation. Hence, the energy density of radiation, produced by gravitational particle production, will eventually dominate the Universe and thus reheat it. As we will show, this is the case in most models we consider in this paper. Interestingly, however, whether kination is realized or not within the non-minimally coupled models we consider depends on the theory of gravity: the so-called metric or Palatini theory. We will make this distinction clear in the following sections.

To summarize, the most important new results obtained in this paper are as follows (i) identification of a non-trivial range of values for the non-minimal coupling function which {\it both} realizes a graceful exit from inflation {\it and} makes the models discussed above consistent with the Planck data, (ii) identification of a suitable reheating mechanism for the above models in scenarios where the usual reheating mechanisms do not work, and (iii) characterization of how the above aspects depend on the theory of gravity (metric or Palatini). As we will show, most of our scenarios naturally include all ingredients of a successful inflationary scenario: a spectral index and tensor-to-scalar ratio consistent with observations, a graceful exit from inflation, and reheating.

The paper is structured as follows. In the next section, we introduce a model of inflation with a non-minimal coupling to gravity and review some basic formulas. In Sec.~\ref{sec:SR_vio}, we discuss how inflation can end by slow-roll violation due to a non-minimal coupling even when the original (minimally coupled) model cannot realize the end of inflation in this way. 
We also show that we can not only invoke slow-roll violation but also make the models we consider viable, i.e., that the predictions for the spectral index and tensor-to-scalar ratio become consistent with the current data, although only for a non-trivial range of the non-minimal coupling value, as we will show. Then in Sec.~\ref{sec:reheating}, we  argue that reheating can be realized via gravitational particle production in most models we consider in this paper. We also briefly discuss the dynamics after inflation. 
The final section is devoted to the summary and conclusions of the paper.

\section{The model} 
\label{sec:setup}

\subsection{Action}

Here we describe our setup to investigate the violation of slow-roll in inflationary models with a non-minimal coupling to gravity. 
The Jordan frame action is assumed as 
\begin{equation}
\label{eq:S_jordan}
	S_{\rm J} 
	= \int d^4x \sqrt{-g}\left(\frac{1}{2} M_{\rm pl}^2 F(\phi)  g^{\mu\nu}R_{\mu\nu}(\Gamma) 
	- \frac{1}{2} g^{\mu\nu}\nabla_{\mu}\phi \nabla_{\nu}\phi - V_J(\phi) \right) \,,
\end{equation}
where $\phi$ is an inflaton and $V_J(\phi)$ is its potential in the Jordan frame, $M_{\rm pl}$ is the reduced Planck mass, $g_{\mu\nu}$ is the metric and $g$ its determinant, $R_{\mu\nu}$ is the Ricci tensor constructed from the space-time connection $\Gamma$ which may or may not depend on the metric and its first derivatives only but also on the inflaton field (see below), and $F(\phi)$ is a function which represents a non-minimal coupling of the inflaton to gravity. In this paper, we assume the following form for this function:
\begin{equation}
\label{eq:F_nonminimal}
F(\phi) =  1+ \xi \left( \frac{\phi}{M_{\rm pl}} \right)^n \,, 
\end{equation}
where $\xi$ is a dimensionless coupling parameter and $n$ is assumed to be a positive integer. 
In this paper, we consider the case with $\xi \ge 0$.
For the potential $V_J(\phi)$ we will discuss
two examples, 
which will be presented
in Secs. \ref{sec:PLI} and \ref{sec:IMI}.
We will also consider two theories of gravity: the so-called {\it metric} and {\it Palatini} theories. In the former case the connection $\Gamma$ depends on the metric only, whereas in the latter case it depends, {\it a priori}, on both the metric and the inflaton field (see Ref. \cite{Bauer:2008zj} for a seminal work and Ref. \cite{Tenkanen:2020dge} for a recent review and introduction to the topic). For simplicity, we will assume that the connection is torsion-free (see, e.g., Refs. \cite{Rasanen:2018ihz,Aoki:2020zqm} for scenarios where this condition was relaxed).

After a Weyl transformation, 
\begin{equation}
\label{eq:weyl_trans}
g_{\mu\nu} \rightarrow \Omega^2 (\phi) g_{\mu\nu}, 
\qquad
\Omega(\phi)^2 \equiv F(\phi) = 1+\xi \left( \frac{\phi}{M_{\rm pl}}\right)^n  \,, 
\end{equation}
the Einstein frame action can be written in both cases as 
\begin{equation}
\label{eq:S_einstein}
	S_{\rm E} 
	= \int d^4x \sqrt{-\hat{g}}\left(\frac{1}{2} M_{\rm pl}^2 \hat{g}^{\mu\nu} \hat{R}_{\mu\nu}(\hat{\Gamma}) 
	- \frac{1}{2} \hat{g}^{\mu\nu} \hat{\nabla}_{\mu}\chi \hat{\nabla}_{\nu}\chi - 
	V_E(\chi) \right) \,,
\end{equation}
where the hat means that the quantity is defined in the Einstein frame and 
where the potential is given by
\begin{equation}
V_E (\chi) = \frac{V_J (\phi(\chi))}{\Omega^4 (\phi(\chi))}  \,.
\end{equation} 
We denote the Einstein frame field by $\chi$, which is related to the Jordan frame counterpart $\phi$ via 
\begin{equation}
\frac{d\phi}{d\chi} 
= \frac{\left(
1+\xi\displaystyle\left(\frac{\phi}{M_{\rm pl}}\right)^n
\right)}
{\sqrt{1+\xi \displaystyle\left(\frac{\phi}{M_{\rm pl}}\right)^n
+\frac32 \kappa n^2 \xi^2 \left( \displaystyle\frac{\phi}{M_{\rm pl}} \right)^{2n-2}}} \,,
\end{equation}
where $\kappa =1\,,0$ correspond to the metric and Palatini cases, respectively.
We can solve  the above equation numerically for an arbitrary $n$ both in the metric and Palatini cases to obtain 
the relation between $\phi$ and $\chi$.  We note that an analytic solution, especially for the $n=2$ case, is well known \cite{GarciaBellido:2008ab,Bauer:2008zj,Rasanen:2017ivk} and, in the Palatini 
case,  the solution even for a general $n$  can be expressed in terms of hypergeometric functions \cite{Jarv:2017azx}. 
In the following, we consider the metric and Palatini cases with $n=4$ for illustrative purposes. 

Once we specify the potential in the Jordan frame, we can calculate the slow-roll parameters and the number of $e$-folds 
by using the Einstein frame potential in the standard fashion. The slow-roll parameters are defined as 
\begin{eqnarray}
\label{eq:SR_param}
	\epsilon &\equiv& \frac{1}{2}M_{\rm pl}^2 \left(\frac{V_E'(\chi)}{V_E(\chi)}\right)^2 \,, \quad
	\eta \equiv M_{\rm pl}^2 \frac{V_E''(\chi)}{V_E(\chi)} \,, 
\end{eqnarray}
where the prime denotes a derivative with respect to $\chi$. Unless some kind of destabilizing mechanism is assumed, inflation ends when slow-roll is violated, $\epsilon(\chi) = 1$. From Eq. \eqref{eq:SR_param} we can also calculate the spectral index $n_s$ and the tensor-to-scalar ratio $r$ as 
\begin{eqnarray}
\label{eq:ns_r}
	n_s  &=& 1 - 6 \epsilon + 2\eta, \qquad
	r =  16 \epsilon \,.
\end{eqnarray}
The spectral index has the measured value $n_s\simeq 0.965$ at the pivot scale $k_*=0.05\, {\rm Mpc}^{-1}$ \cite{Akrami:2018odb}, whereas the tensor-to-scalar ratio is constrained to $r<0.06$ \cite{Ade:2018gkx}.

In this paper, we consider two types of inflation models and show that while their minimally coupled versions predict values of $n_s$ and $r$ that are excluded by the data, their non-minimally coupled extensions can be easily resurrected. We will describe these models in more detail in the following subsections.

\subsection{Inflation models}

\subsubsection{Power-law inflation}
\label{sec:PLI}

To facilitate comparison with the minimally coupled case, we give the inflaton potentials in the Jordan frame. For the power-law inflation \cite{Abbott:1984fp}, the potential is given by
\begin{equation}
V_J(\phi) = V_0 e^{-\alpha \phi/M_{\rm pl}}\,,
\end{equation}
where $\alpha$ is a dimensionless parameter and $V_0$ is a parameter representing a scale which is roughly the same as the energy scale of inflation.
A potential like this can arise in supergravity and string theories,  and in some models 
a successful inflationary scenario with $\alpha \ll 1$ can be constructed,
for example, in the framework of M theory \cite{Becker:2005sg}.

In the minimally coupled case, the slow-roll parameters are given by 
\begin{equation}
\label{eq:SR_PLI}
\epsilon = \frac12 \alpha^2, 
\qquad 
\eta = \alpha^2 \,.
\end{equation}
Since $\alpha$ is assumed to be a constant, the slow-roll parameters in this model are also constants. 
Therefore, inflation cannot end by violation of slow-roll caused by the dynamics of the inflaton. Therefore, in this model, one needs a non-standard mechanism to end inflation, such as tachyonic instability (see, e.g., Ref. \cite{Martin:2013tda} for details).

The need for an extra mechanism to end inflation is not the only problem of this model. From Eqs.~\eqref{eq:ns_r} and~\eqref{eq:SR_PLI} one can derive a relation between $n_s$ and $r$: 
\begin{equation}
r = -8 (n_s -1).
\end{equation}
Since recent observations imply $n_s \simeq 0.965$, the above relation indicates that $r \simeq 0.28$, which is excluded by observations \cite{Akrami:2018odb,Ade:2018gkx}. However, as we will see in the next section, by introducing a non-minimal coupling, the slow-roll parameters can evolve in time and, consequently, violation of slow-roll can be invoked. Furthermore, the predictions for the spectral index and the tensor-to-scalar ratio will also get modified, and the tension with the data can be alleviated for a sufficient choice of the non-minimal coupling function 
that depends on the $\alpha$ parameter in the potential.

\subsubsection{Inverse monimial inflation} 
\label{sec:IMI}

The Jordan frame potential of the inverse monomial inflation model is given by \cite{Peebles:1987ek,Ratra:1987rm}
\begin{equation}
V_J(\phi) = V_0 \left(\frac{\phi}{M_{\rm pl}}\right)^{-p} ,
\end{equation}
where $p$ is a positive number and $V_0$ represents an energy scale. Models with an inverse monomial potential have been discussed in the context of, e.g., quintessential inflation \cite{Peebles:1987ek,Ratra:1987rm}, intermediate inflation \cite{Barrow:1993zq},
tachyon inflation \cite{Feinstein:2002aj,Sami:2002zy}, and 
dynamical supersymmetric inflation \cite{Kinney:1997hm,Kinney:1998dv}.

In the minimally coupled case, the slow-roll parameters are
\begin{equation}
\epsilon = \frac12 p^2 \left( \frac{\phi}{M_{\rm pl}} \right)^{-2},
\qquad
\eta =  p(p+1) \left( \frac{\phi}{M_{\rm pl}}\right)^{-2} \,,
\label{eq:slow_roll_IMI}
\end{equation}
from which one obtains
\begin{equation}
n_s - 1 = p (2-p) \left( \frac{\phi}{M_{\rm pl}}\right)^{-2}~.
\end{equation}
From this expression, one can see that the spectral index is blue-tilted when $p <2$, which is excluded by observations. On the other hand, 
Eq.~\eqref{eq:slow_roll_IMI} indicates the relation
\begin{equation}
r = \frac{8p}{2-p} (n_s -1 ) \,,
\end{equation}
from which one can see that even when $p > 2$ 
and $n_s = 0.965$, the tensor-to-scalar ratio is predicted as $r > 0.28$, i.e. it is bounded from below, whereas observations indicate $r<0.06$ \cite{Ade:2018gkx}. Therefore, the minimally coupled version of this model is completely excluded by observations. 

Furthermore, since in the minimally coupled case inflation starts at small values of $\phi$ and the field value grows during inflation, the slow-roll parameters monotonically decrease as inflation proceeds. Therefore, also in this model inflation cannot end by violation of slow-roll driven by the inflationary dynamics without an additional mechanism.

We will see that by introducing a non-minimal coupling, we can realize a graceful exit from inflation in this model and obtain values of the spectral index and tensor-to-scalar ratio consistent with the current data at the same time.

\section{Violation of slow-roll and observables in the non-minimally coupled case} 
\label{sec:SR_vio}

In this section, we consider the inflation models mentioned in the previous section but this time in the non-minimally coupled case and show how slow-roll can be violated by the existence of a non-minimal coupling to gravity, which ends inflation.\footnote{For some early, pre-Planck works on the topic in a somewhat different context, see Refs.~\cite{Frewin:1993aj,Kaganovich:2000fc}.}
We will also study predictions for the observables $n_s$ and $r$ in this context.\footnote{In addition to providing results for these observables, the Planck data also constrain the running and running of the running of the spectral index \cite{Akrami:2018odb}. We have checked that in all cases we present, the running parameters are small and well within the limits given by the observations.} We present our results for each model in order.

\subsection{Non-minimal power-law inflation}
\label{sec:nm_PLI}

First we discuss the case of the power-law inflation (PLI) model. For illustrative purposes, as mentioned in Sec.~\ref{sec:setup}, we assume $n=4$ for the non-minimal coupling function, i.e.,
\begin{equation}
\Omega^2 (\phi) = 1 +\xi \left( \frac{\phi}{M_{\rm pl}}\right)^4 \,.
\end{equation}

\begin{figure}
\begin{center}
\includegraphics[width=9cm]{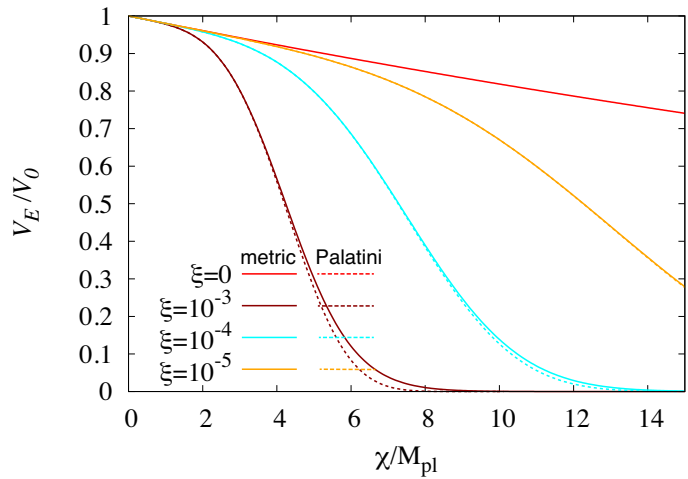}
\caption{Potential for the power-law inflation model with $\alpha=0.02$ for the cases $\xi=0, 10^{-3}, 10^{-4}$ and $10^{-5}$.
The metric and Palatini cases are shown. 
}
\label{fig:PLI_potential}
\end{center}
\end{figure}

The Einstein frame potential is depicted in Fig.~\ref{fig:PLI_potential} for the cases of $\xi = 0$ 
(minimally coupled case), $10^{-3}, 10^{-4}$, and $10^{-5}$. In Fig.~\ref{fig:PLI_potential}, we take $\alpha = 0.02$ as a representative example. As can be seen from the figure, as $\xi$ increases, the potential becomes more steep, which causes the slow-roll parameters to increase as the field evolves. This is not surprising, as the non-minimal coupling changes the Einstein frame potential. However, when $\alpha \sim 0.02$, for $\xi \gtrsim 10^{-3}$ the potential becomes too steep to support more than roughly $50-60$ $e$-folds, and therefore for a too large non-minimal coupling it becomes questionable whether the non-minimal PLI model can solve the classic horizon and flatness problems 
(see, e.g., Ref. \cite{Liddle:2003as}). For this reason, we only show results in the PLI case for $0 \leq \xi \leq 10^{-3}$ in this paper.
It should be emphasized that whether a model of inflation can be made viable needs to be investigated carefully since the introduction of a non-minimal coupling does not necessarily guarantee the success of the model.
One needs to assume a suitable value for $\xi$ depending on the model and its parameters.

\begin{figure}
\begin{center}
\includegraphics[width=7.5cm]{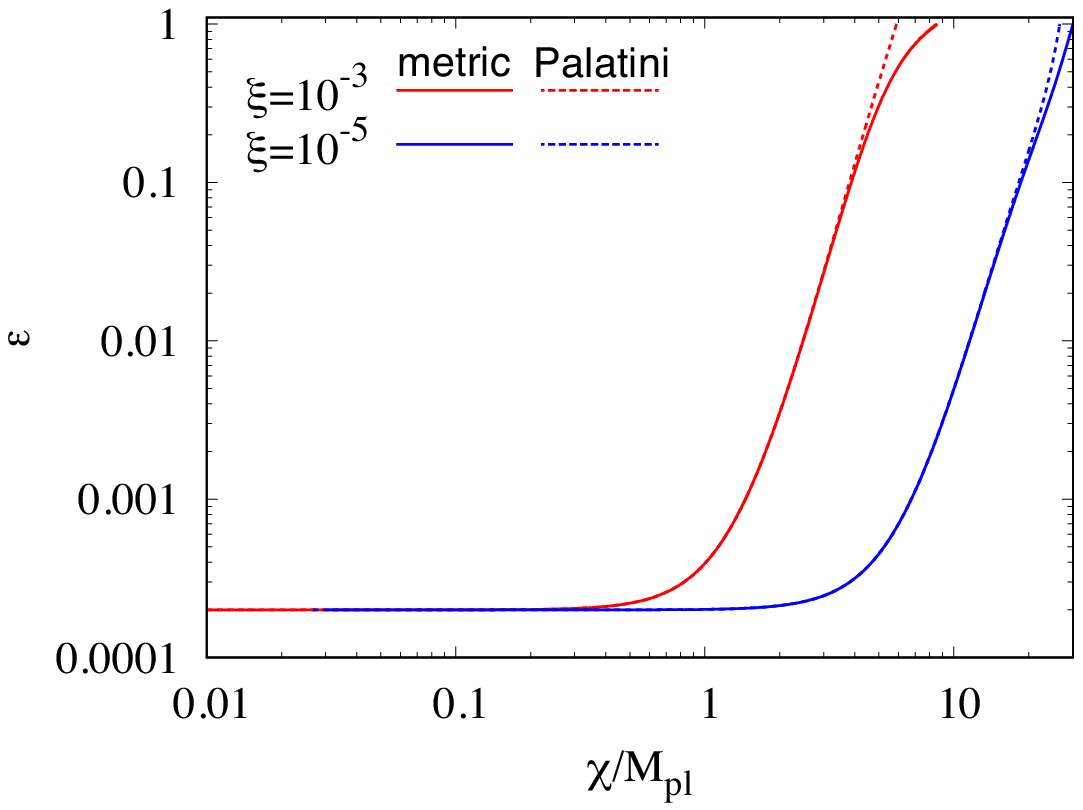}
\includegraphics[width=7.5cm]{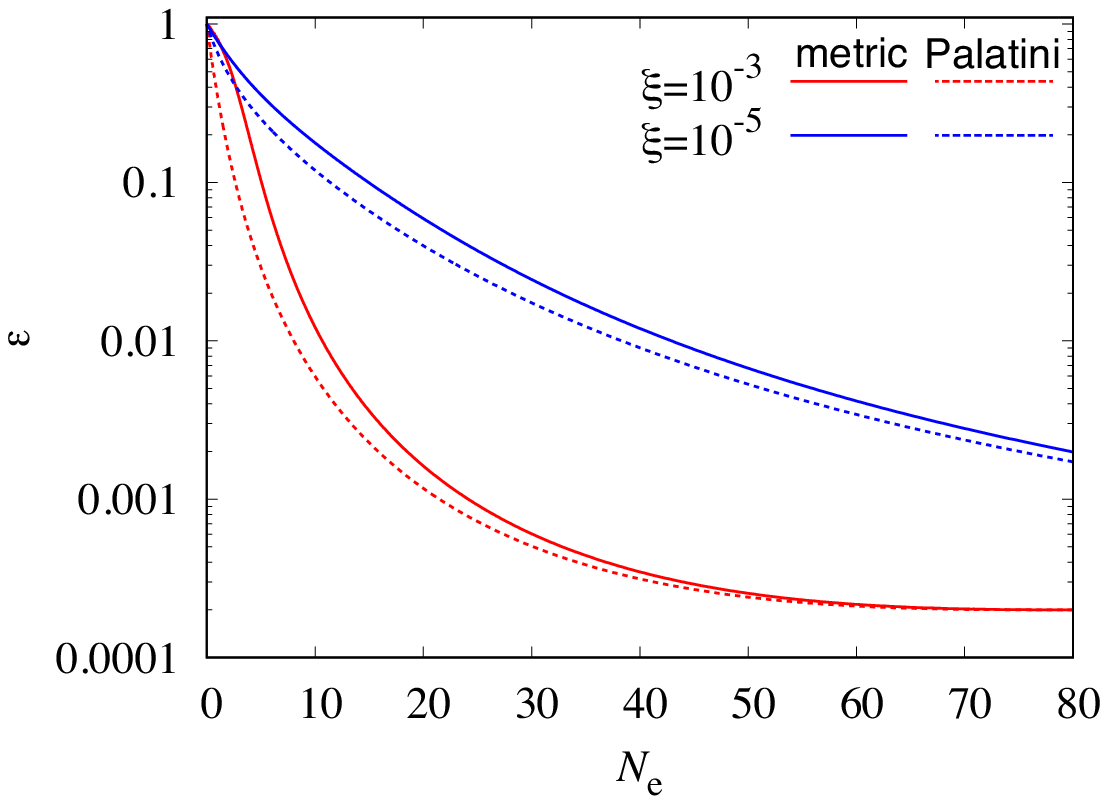}
\caption{Evolution of $\epsilon$ as a function of $\chi$ (left) and $N_e$ (right) in the power-law inflation model with $\alpha=0.02$ for the cases $\xi=10^{-3}$ and $\xi=10^{-5}$.
The metric and Palatini cases are shown. 
}
\label{fig:PLI_epsilon}
\end{center}
\end{figure}

In the left panel of Fig.~\ref{fig:PLI_epsilon}, the evolution of one of the slow-roll parameters, $\epsilon$, is shown for the case with $\alpha=0.02$. Inflation starts from small values of $\chi$ and, as inflation proceeds, the field evolves down the potential towards a larger value. 
For $\alpha = 0.02$, the minimally coupled case gives $\epsilon = 2 \times 10^{-4}$, and hence even in the non-minimally coupled case, $\epsilon$ starts from 
the value corresponding to the minimally coupled case.
As $\chi$ increases
and the non-minimal coupling term becomes larger,
$\epsilon$ gets larger, as can be read off from the figure.
 One can also notice that as $\xi$ becomes larger, $\epsilon$ increases to reach unity at smaller values of $\chi$; i.e., the violation of slow-roll occurs faster for larger values of $\xi$.

\begin{figure}[h]
\begin{center}
\includegraphics[width=12cm]{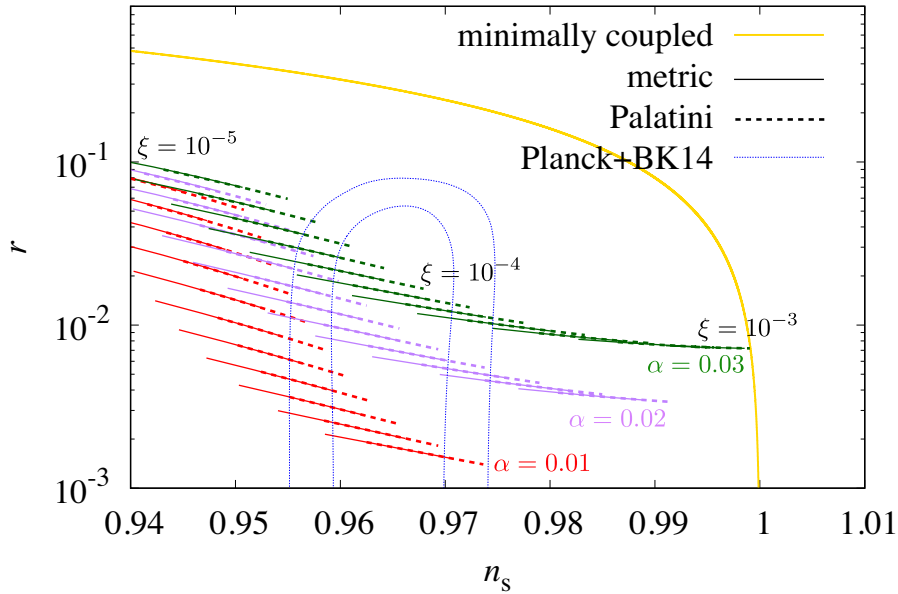}
\caption{Predictions for the spectral index $n_s$ and tensor-to-scalar ratio $r$ in the non-minimally coupled power-law inflation model with $\alpha=0.01, 0.02$ and $0.03$. We vary the non-minimal coupling parameter as $ 10^{-5} \leq \xi \leq 10^{-3}$. The yellow curve indicates the predictions in the minimally coupled case, and the underlying blue regions indicate the Planck+BICEP2/Keck Array $1\,\sigma$ and $2\,\sigma$ bounds \cite{Array:2015xqh,Akrami:2018odb}. The number of $e$-folds is assumed as $50 \leq N_e \leq 60$ in this figure. The metric and Palatini cases are shown. 
}
\label{fig:PLI_ns_r}
\end{center}
\end{figure}

In the right panel of Fig.~\ref{fig:PLI_epsilon}, the evolution of $\epsilon$ is plotted as a function of $N_e$, the number of $e$-folds counted backwards from the end of inflation. The figure again illustrates that 
when $\xi$ is large, $\epsilon$ goes up relatively quickly. However, 
as one can see in Fig.~\ref{fig:PLI_ns_r}, if $\xi$ is too large, the predictions for $n_s$ and $r$ get close to the minimally coupled case even though slow-roll violation is quickly realized.
In Appendix \ref{app:SRP_behavior}, we discuss the evolution of the slow-roll parameter in more detail, focusing on the dependence on $\xi$, which helps us understand the dynamics
and the non-trivial $\xi$-dependence of $n_s$ and $r$.

The above aspects indicate that for a successful inflation model, we need (mild) modifications around $N_e \sim 50 - 60$ to obtain predictions for the spectral index and tensor-to-scalar ratio which are consistent with observations. 
These predictions are shown in Fig.~\ref{fig:PLI_ns_r} in the slow-roll approximation. In the figure, we take $\alpha=0.01, 0.02$,  and $0.03$ for illustrative purposes. For larger~$\alpha$, the tensor-to-scalar ratio~$r$  gets larger. 
As can be seen from the right panel of Fig.~\ref{fig:PLI_epsilon}, the value of $\epsilon$ around $N_e \sim 50 -60$ for the case with a relatively large $\xi$ is close to that in the minimally coupled model, 
which means that the prediction for $r$  approaches $r =16\epsilon \sim 8 \alpha^2$. Therefore, when $\alpha \gtrsim 0.1$, the tensor-to-scalar ratio becomes $r \gtrsim 0.08$ regardless of $\xi$, which is not consistent with observations even with a non-minimal coupling to gravity. Figure~\ref{fig:PLI_ns_r} also shows that the differences between the metric and Palatini cases are rather modest, which is due to $\xi$ taking a value much smaller than unity. This is reminiscent of the behavior found in, e.g., Refs. \cite{Tenkanen:2017jih,Racioppi:2017spw,Jarv:2017azx}, which also studied inflationary models with a small non-minimal coupling to gravity in both the metric and Palatini frameworks.  Finally, we note that the 
parameter space of the non-minimal PLI model presented in Fig.~\ref{fig:PLI_ns_r} 
is testable with forthcoming CMB missions. For example, future CMB B-mode polarization experiments
such as BICEP3 \cite{Wu:2016hul}, LiteBIRD~\cite{Matsumura:2013aja}, and the Simons Observatory \cite{Simons_Observatory} will be soon pushing the limit on tensor-to-scalar ratio down to $r\simeq 0.001$,  
or aiming to detect $r$ above this limit. These measurements will either provide further support for the model or rule out a large part of its parameter space. For a discussion on prospects for distinguishing between different non-minimal models in the case of a detection, see Ref. \cite{Takahashi:2018brt}.

\subsection{Non-minimal inverse monomial inflation}

\begin{figure}
\begin{center}
\includegraphics[width=9cm]{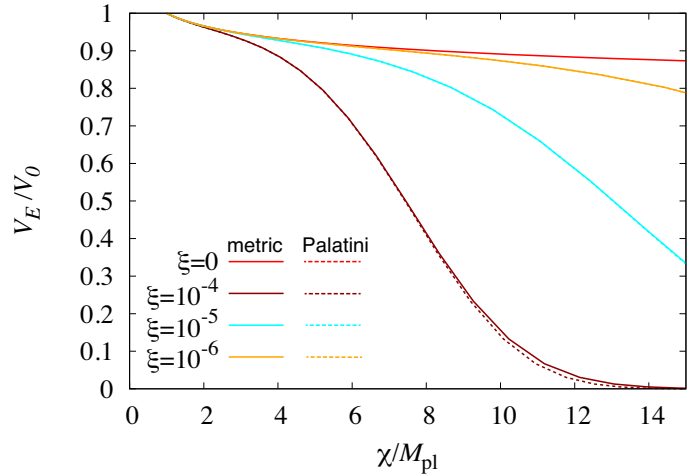}
\caption{Potential for the inverse monomial inflation 
model with $p=0.05$ in the cases $\xi=0,10^{-4}, 10^{-5}$ and $\xi=10^{-6}$.
 The metric and Palatini cases are shown. 
 }
\label{fig:IMI_potential}
\end{center}
\end{figure}

\begin{figure}[h]
\begin{center}
\includegraphics[width=12cm]{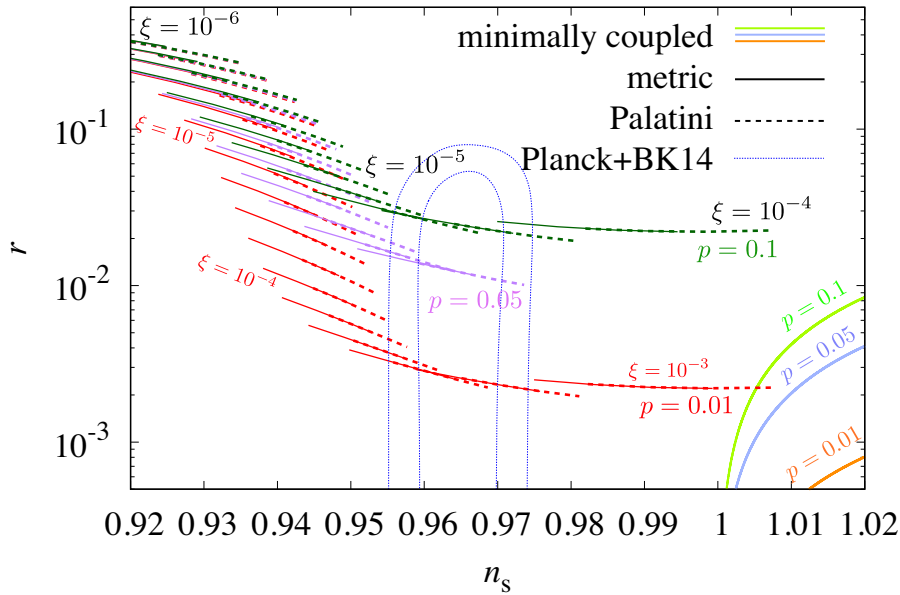}
\caption{Predictions for the spectral index $n_s$ and tensor-to-scalar ratio $r$ in the non-minimally coupled inverse monomial inflation 
model with $p=0.01,\, 0.05$, and $0.1$. For $p=0.05,\, 0.1$, we vary the non-minimal coupling parameter as $ 10^{-6} \leq \xi \leq 10^{-4}$, whereas for $p=0.01$ we vary it between $ 10^{-6} \leq \xi \leq 10^{-3}$ because in this case the potential can support more than $60$ $e$-folds of inflation also for $\xi> 10^{-4}$.
The predictions of the minimally coupled case are also shown and the underlying blue regions indicate the Planck+BICEP2/Keck Array $1\, \sigma$ and $2\,\sigma$ bounds \cite{Array:2015xqh,Akrami:2018odb}. The number of $e$-folds is assumed as $50 \leq N_e \leq 60$ in this figure. The metric and Palatini cases are shown. 
}
\label{fig:IMI_ns_r}
\end{center}
\end{figure}

Let us now discuss the inverse monomial inflation (IMI) model in the framework of non-minimal coupling to gravity. 
Here we again assume $n=4$ for the non-minimal coupling function. In Fig.~\ref{fig:IMI_potential}, we show some example Einstein frame potentials in the case of non-minimal version of the model with $p=0.05$
and $\xi=10^{-4}, 10^{-5}$ and $10^{-6}$, as well as with $\xi = 0$.
As in the PLI case, the inflaton moves from a small value to a large one during inflation. 
As can be seen from Fig.~\ref{fig:IMI_potential}, a non-minimal coupling again makes the potential steeper, which drives the $\epsilon$ parameter to larger values and eventually ends inflation, in contrast to the minimally coupled counterpart of this model. Similar to the non-minimal PLI model discussed in Sec.~\ref{sec:nm_PLI}, also in this case the potential becomes too steep to support more than roughly $60$ $e$-folds\footnote{While this value is compatible with what is shown in Fig.~\ref{fig:IMI_ns_r}, it is not enough to solve the classic horizon and flatness problems in a scenario where the scale of inflation is high and inflation is followed by a 
``kination" phase \cite{Liddle:2003as}. We will return to the dynamics after inflation in Sec. \ref{sec:reheating}.} if the non-minimal coupling is larger than $\xi \gtrsim 10^{-4}$ 
for $p\sim 0.05$. For smaller $p$, however, the potential can also support more than $60$ $e$-folds for larger $\xi$.

\begin{figure}
\begin{center}
\includegraphics[width=7.5cm]{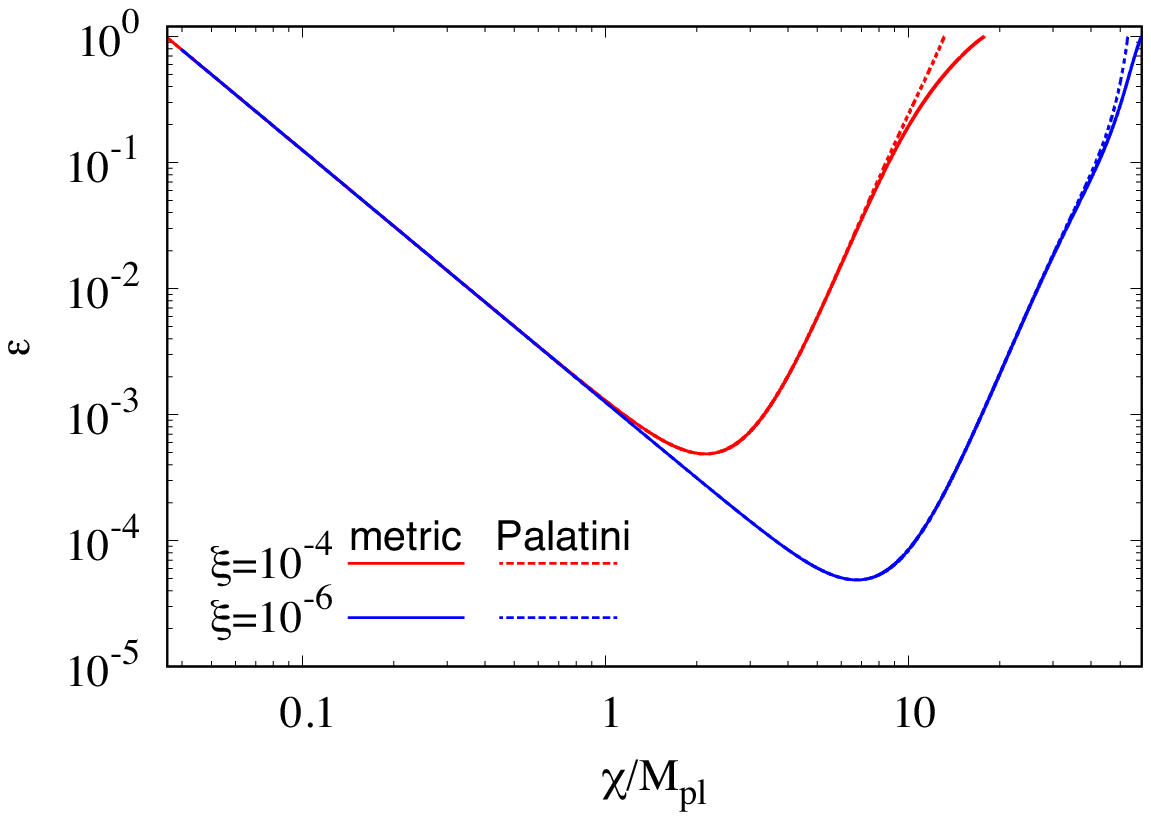}
\includegraphics[width=7.5cm]{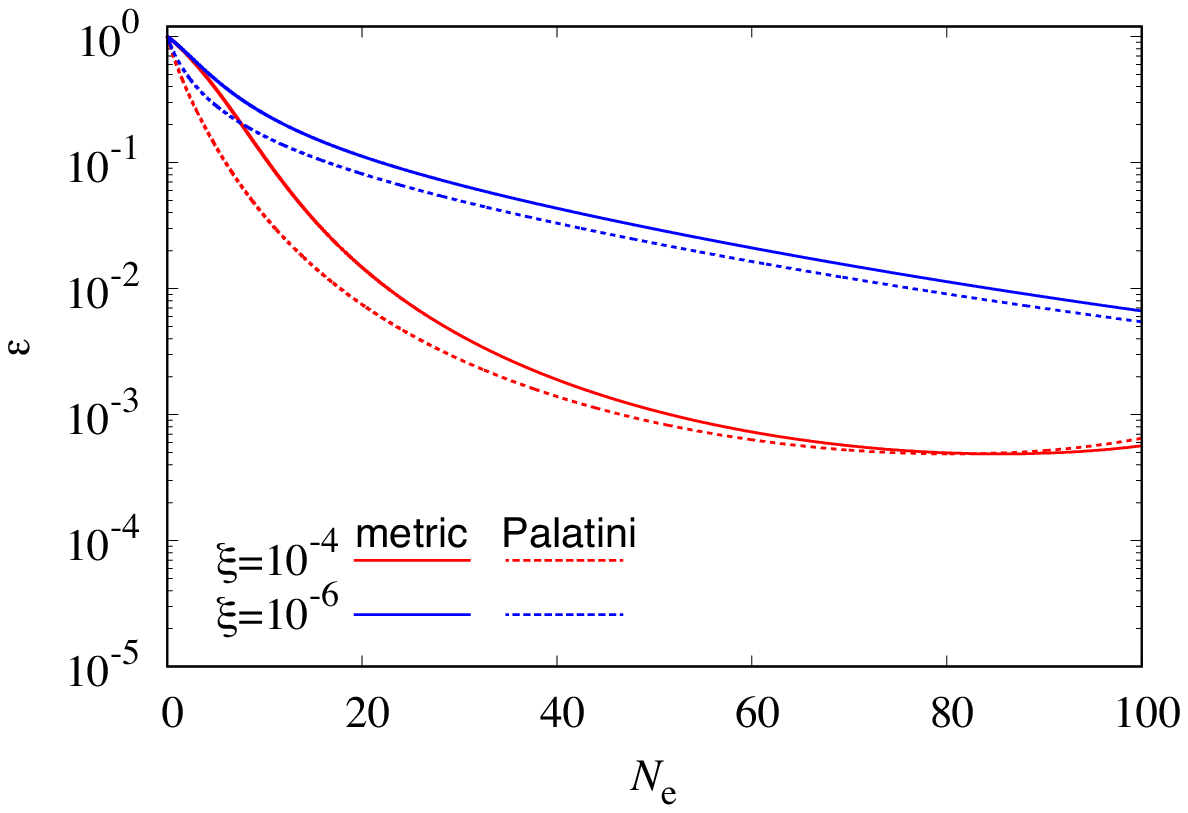}
\caption{Evolution of $\epsilon$ as a function of $\chi$ (left) and $N_e$ (right) in the inverse monomial inflation  
model with $p=0.05$ in the cases $\xi=10^{-4}$ and $\xi=10^{-6}$. The metric and Palatini cases are shown. 
}
\label{fig:IMI_epsilon}
\end{center}
\end{figure}

In Fig.~\ref{fig:IMI_epsilon}, the evolution of $\epsilon$ as a function of $\chi$ (left panel) and $N_e$ (right panel) is shown for the example cases $\xi =10^{-4}$ and $\xi = 10^{-6}$. We  take $p=0.05$ in this figure too.  Although the tendency is the same as in the case of the PLI model, in the case of the IMI model, $\epsilon$ first gets smaller during the early 
stages of inflation, which is a characteristic of the minimally coupled model. However,
as $\chi$ grows, the non-minimal coupling term becomes more dominant and $\epsilon$ becomes larger. Then,  it finally ends inflation, which again is the effect of the non-minimal coupling.

In Fig.~\ref{fig:IMI_ns_r}, the predictions of the IMI model for $n_s$ and $r$ in the non-minimally coupled case are shown in slow-roll approximation for $p=0.01, 0.05$ and $0.1$.
We take the number of $e$-folds as $50 \leq N_e \leq 60$. 
Regarding the non-minimal coupling parameter, we vary it as $10^{-6} \leq \xi \leq 10^{-3}$ for $p=0.01$, but for $p=0.05$ and $0.1$,  we take a narrower range  $10^{-6} \leq \xi \leq 10^{-4}$
since the potential cannot support more than $60$ $e$-folds for $\xi \ge 10^{-4}$ in these cases.
These are interesting values of $p$, as they give a blue-tilted $n_s$ in the original minimally coupled case and hence are totally excluded by observations.  However, due to the existence of the non-minimal coupling, the (Einstein frame) potential gets modified and the spectral index can become red-tilted so that the model is viable for some range of $\xi$ again, just as in the case of the PLI model. 
If we take $p$ larger than $p=0.5$, the spectral index can still become negative due to the existence of the non-minimal coupling but, on the other hand, the tensor-to-scalar ratio gets as large as $r\sim 0.1$. Therefore, models with $p \gtrsim 0.5$ cannot be made viable even with the existence of a non-minimal coupling of the type studied in this paper.  Finally, similar to the non-minimal PLI model, Fig.~\ref{fig:IMI_ns_r} also shows that the differences between the metric and Palatini cases are small due to the non-minimal coupling taking only small values. Also, similar to the non-minimal PLI model, forthcoming CMB B-mode polarization experiments 
will soon test the model, in particular, the parameter space presented in Fig.~\ref{fig:IMI_ns_r}.

\section{Reheating and dynamics after inflation}
\label{sec:reheating}

As shown in the previous section, due to the existence of the non-minimal coupling,
we can dynamically realize a graceful exit in both the PLI and IMI models; that is, we can obtain $\epsilon = 1$ without any additional mechanism.
In addition to having a successful mechanism for ending inflation, in a successful inflationary model the Universe must also be reheated so that by the time of big bang nucleosynthesis, the Universe becomes radiation dominated. 

In the standard reheating scenario (see, e.g., Ref.~\cite{Kofman:1997yn}), the inflaton field starts to oscillate around its potential minimum after the end of inflation and the energy density of the oscillating scalar field evolves roughly in the same way as that of non-relativistic matter. During such an effectively matter-dominated epoch, the inflaton field can decay into radiation through some interaction and, as a result, the Universe can 
be reheated.
On the other hand, the models discussed here, as seen in Figs.~\ref{fig:PLI_potential} and \ref{fig:IMI_potential}, do not have any potential minimum even with a non-minimal coupling, and hence we cannot expect the usual reheating mechanism described above to work in our setup. However, the so-called gravitational reheating can still be realized in most of our scenarios, as we will explain below.

\begin{figure}
\begin{center}
\includegraphics[width=7.5cm]{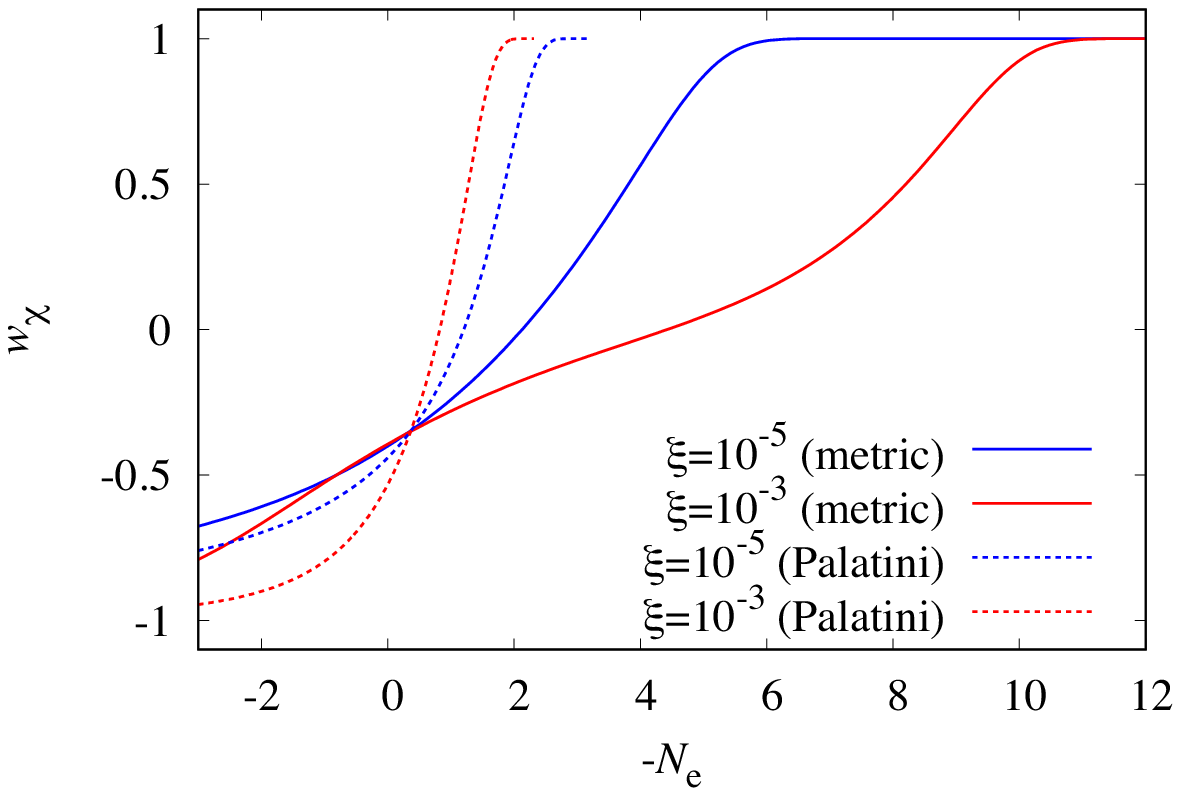}
\includegraphics[width=7.5cm]{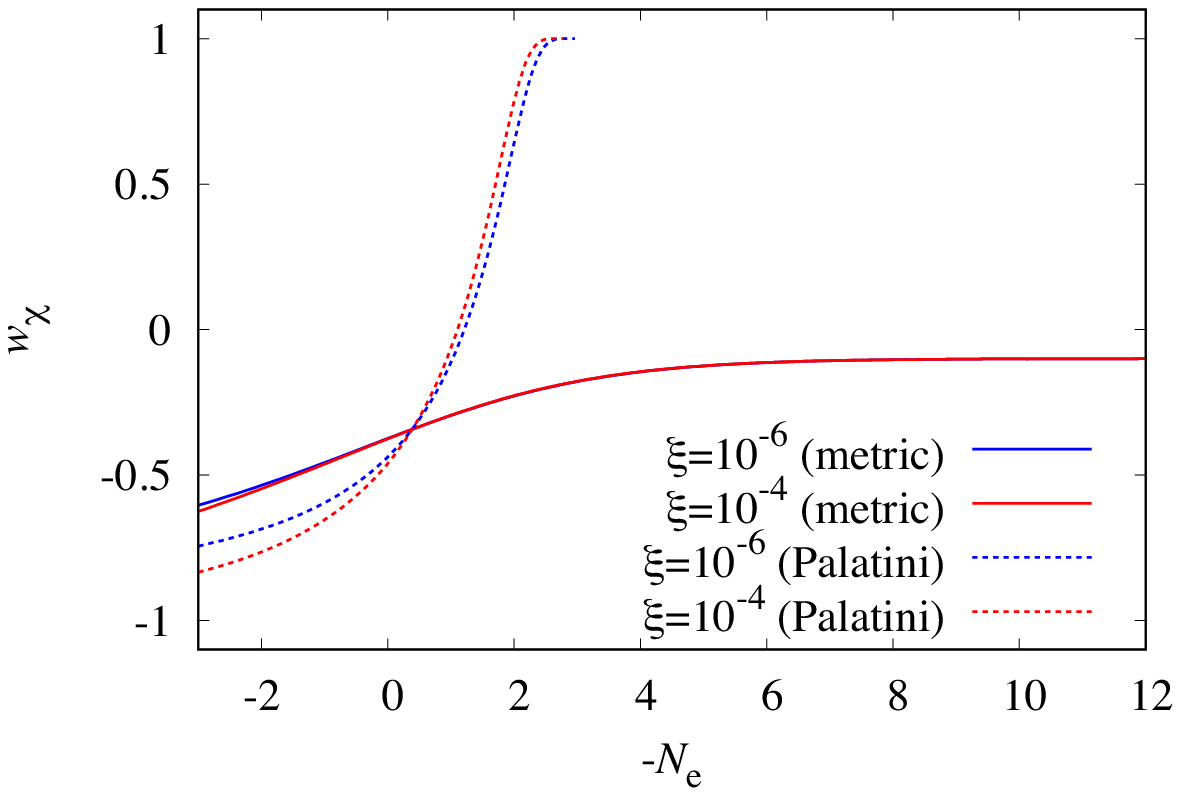}
\caption{Evolution of the equation-of-state parameter $w$ as a function of the number of $e$-folds measured from the end of inflation. Left panel: The power-law inflation model with $\alpha =0.02$ in the cases of $\xi=10^{-3}$ and $\xi=10^{-5}$. 
Right panel: The inverse monomial inflation model with $p=0.05$ in the cases of $\xi=10^{-4}$ and $\xi=10^{-6}$.  In both panels, the metric and Palatini cases are shown.
}
\label{fig:PLI_eos}
\end{center}
\end{figure}

The effective equation-of-state parameter of the inflaton field in the Einstein frame, $w_\chi$, is defined by
\begin{equation}
    w_\chi \equiv \frac{ \dot{\chi}^2 /2 - V_E(\chi)}{ \dot{\chi}^2 /2 + V_E(\chi)}~.
\end{equation}
In Fig.~\ref{fig:PLI_eos}, we plot the evolution of this quantity in the PLI and IMI models as a function of the number of $e$-folds measured from the end of inflation; that is, $N_e=0$ in the figure corresponds to the end of inflation and $-N_e>0$ to the post-inflationary stage. As can be seen in the  left panel of the figure, in the  PLI model considered here the equation-of -state parameter reaches unity soon after the end of inflation, both in the Palatini and in the metric case. This $w_\chi=1$ phase is nothing but the so-called kination, where the kinetic energy of the field dominates the total energy density of the Universe. However, as can be seen in the right panel of Fig. \ref{fig:PLI_eos}, only in the Palatini counterpart of the IMI model is it possible to reach $w_\chi=1$ and reheat the Universe via gravitational production of particles.
In the metric case with $p\sim 0.05$, the potential is not steep enough to give $w_\chi=1$, but instead we obtain $w_\chi \sim -0.1$.  The reason for this unusual behavior is explained in Appendix \ref{app:IMI_eos}.

Reheating mechanisms in scenarios where the Universe undergoes a kination epoch have been studied in a large number of works. One possibility is gravitational reheating, which is based on gravitational particle production that
{\it necessarily} occurs due to excitation of all light fields during or at the end of  inflation \cite{Ford,Starobinsky:1994bd}. 
It has been shown that in this way, and under some suitable conditions regarding the non-minimal coupling between the Standard Model (SM) Higgs field and gravity, even the SM Higgs can be excited either during or after inflation; thus, it becomes responsible for reheating after inflation regardless of the exact shape of the potential or couplings of the inflaton field \cite{Figueroa:2016dsc,Nakama:2018gll,Opferkuch:2019zbd}. For other recent studies on this mechanism, see Refs. \cite{Dimopoulos:2018wfg,Haro:2018jtb,Hashiba:2018iff,Bettoni:2019dcw}. For earlier studies on gravitational reheating in the context of non-minimal inflation, see Refs. \cite{Tashiro:2003qp,Watanabe:2006ku}.

Let us now analyze how long it takes for the Universe to get reheated. While the energy density of radiation produced gravitationally is at first only quite modest, $\rho_{\rm rad} \sim H_*^4 \ll H_*^2M_{\rm pl}^2 \sim \rho_{\rm tot}$, where $H_*$ is the Hubble scale and $\rho_{\rm tot}\sim \rho_{\chi}$ is the total energy density at the end of inflation,\footnote{More precisely, depending on how the transition from inflation to the kination epoch proceeds, the efficiency of the gravitational particle production is reduced and the energy density is given by
$\rho_{\rm rad}   = {\cal A} H_*^4$ where ${\cal A} =  {\cal O}(0.01) - {\cal O}(0.1) $ \cite{Chun:2009yu}. However, even when ${\cal A} = 0.01$, 
the
qualitative picture does not change.} 
the radiation component's energy density 
scales down more slowly than that of the inflaton field as the Universe expands, which eventually reheats the Universe. This is indeed the case when the inflaton enters into a kination phase where its kinetic energy dominates the energy density, $\rho_\chi \sim \dot{\chi}^2/2 \gg V(\chi)$ and consequently $\rho_\chi \propto a^{-6}$, where $a$ is the scale factor. For radiation, $\rho_{\rm rad} \propto a^{-4}$ as usual, and therefore $\rho_{\rm rad}/\rho_\chi \propto a^2$. Thus, the Universe becomes radiation dominated in 
\begin{equation}
\label{reheating_efolds}
N_{\rm reh} \simeq \ln\left(\frac{M_{\rm P}}{H_*}\right) \simeq 10 - 12\,,
\end{equation}
$e$-folds after the end of inflation. Here we used the definition of the tensor-to-scalar ratio
\begin{equation}
r \equiv \frac{\mathcal{P}_T}{\mathcal{P}_\zeta} = \frac{8}{M_{\rm P}^2 \mathcal{P}_\zeta}\left(\frac{H_*}{2\pi}\right)^2\,,
\end{equation}
which allows us to estimate
\begin{equation}
\label{Hk}
H_* \simeq 7.7\times 10^{13}\sqrt{\frac{r}{0.1}}\, {\rm GeV} \simeq  7.7\times \left(10^{12} - 10^{13}\right) {\rm GeV} \,,
\end{equation}
as in our scenarios $r\sim 0.001 - 0.1$ and the curvature power spectrum amplitude $\mathcal{P}_\zeta = 2.1\times 10^{-9}$ as given by the Planck observations \cite{Akrami:2018odb}. The time of reheating \eqref{reheating_efolds} corresponds to the reheat temperature\footnote{A more detailed analysis of the reheating temperature in the gravitational reheating scenario has been done in Ref.~\cite{Hashiba:2018iff}.}
\begin{equation}
T_{\rm reh} \simeq \rho_{\rm rad}^{1/4}(N_{\rm reh}) \simeq H_* e^{-N_{\rm reh}} \simeq \left(10^7 - 10^9\right)\, {\rm GeV}\,,
\end{equation}
which is well above the temperature required for successful BBN, $T_{\rm BBN} = \mathcal{O}(1)$ MeV (see, e.g., Ref. \cite{Hasegawa:2019jsa}). We therefore conclude that gravitational particle production is sufficient to reheat the Universe in a successful way.

We stress that in most of our scenarios, the conditions for a successful inflationary model can be satisfied due to the existence of the non-minimal coupling even though the minimally coupled versions cannot accommodate a graceful exit or reheating without some extra mechanisms. However,  there is still a serious issue in this simple setup: large running of the gravitational coupling in what we call the Jordan frame, provided that standard matter is minimally coupled in that frame \cite{Carroll:1998zi,Dvali:2001dd,Chiba:2001er}. In order to solve this problem, one may need to further assume that, for example, the standard matter is minimally-coupled in the Einstein frame (instead of the Jordan frame) or that the inflaton field has a coupling which changes its potential at large field values and/or eventually stops rolling after inflation. However, this issue is beyond the scope of the present paper and we leave it for future work.

\section{Conclusions} 
\label{sec:summary}

In this paper, we have shown that a non-minimal coupling to gravity can not only make some inflationary models viable, similar to the case of the Higgs inflation model, but can also invoke slow-roll violation to realize a graceful exit from inflation. In particular, this is the case for models where some destabilizing mechanism, such as tachyonic instability, should be assumed to end inflation when the 
model is minimally coupled to gravity.

As explicit examples, we have considered the power-law and inverse monomial inflation models with a non-minimal coupling to gravity.
When coupled only minimally to gravity, these models are completely excluded since their predictions for the spectral index $n_s$ and tensor-to-scalar ratio $r$ are inconsistent with the current cosmological data. However, we have shown that these models can become viable again with a specific range of values of the non-minimal coupling to gravity. 
Typical values are summarized in Table~\ref{table:summary}.
\begin{table}[]
\begin{tabular}{cccc} \hline \hline
PLI ($\alpha =$) & 0.01 & 0.02 & 0.03 \\ \hline
$ \xi $   & $ 1 \times 10^{-3}$ & $ 4 \times 10^{-4}$  & $ 1 \times 10^{-4}$   \\ \hline \hline 
IMI ($p =$) & 0.01 & 0.05 & 0.1 \\ \hline
$ \xi  $ & $ 7 \times 10^{-4}$  & $ 1 \times 10^{-4}$  & $ 7 \times 10^{-5}$  \\ \hline
\end{tabular}
\caption{A summary of typical values of the non-minimal coupling to gravity,
which can make the power-law and inverse monomial inflation models viable again.
Note that here we assume $n=4$ for the non-minimal coupling function.}
\label{table:summary}
\end{table}
The characterization of this range is  one of our most important results. 

Furthermore, the same non-minimal coupling can also invoke the required slow-roll violation to end inflation without the need to implement any other mechanism for a graceful exit; this is non-trivial since introducing a non-minimal coupling does not necessarily guarantee that either the modifications to $n_s$ and $r$ are consistent with the Planck data or that the slow-roll violation can be realized. We also showed that in both the PLI and IMI models considered in this paper, the forthcoming CMB B-mode polarization experiments will soon either provide further support for the models or rule out a large part of their parameter space. 

Our findings facilitate model building of the inflationary Universe, especially in the framework with an extended gravity sector. In this paper we have studied both metric and Palatini theories of gravity. We found that when it comes to inflationary observables, the differences between the two theories are generically small in the models discussed in this paper. This is due to the non-trivial fact that in our scenarios compatibility with data requires the non-minimal coupling to gravity be very small, $\xi \ll 1$, in contrast to many other models such as Higgs inflation, which require very large non-minimal couplings in order to be compatible with the data. However, as we also showed, the post-inflationary dynamics of the inflaton field can -- surprisingly -- be drastically different in the two counterparts of the same model, depending on the underlying theory of gravity. 

The above notion has important consequences for reheating. Since there is no potential minimum in our setup, the usual reheating mechanism where the oscillating inflaton field decays into radiation 
cannot work. However, in most of our scenarios a kination phase is typically realized just after the end of inflation, which allows gravitational  particle production to complete reheating at temperatures well above those required for successful BBN. 
Among the models we considered in this paper,  the only exception is the non-minimally coupled inverse monomial inflation model within the metric theory of gravity. This highlights the fact that even when the differences between two theories of gravity are small as far as inflationary observables are concerned, their suitability for building a successful model of inflation can be dramatically different. However, as shown in the paper, most of our models can accommodate all three major ingredients of a successful inflationary model: predictions for $n_s$ and $r$ consistent with data, a graceful exit, and reheating. However, in all scenarios there still remains an issue regarding the running of the gravitational constant at late times, as discussed briefly at the end of the previous section. A detailed study of this issue is left for future work.

\acknowledgments
S.~Yokoyama would like to thank Soichiro Hashiba for useful discussions.
The work of T.~Takahashi was supported by JSPS KAKENHI Grant Number 17H01131,  19K03874 and MEXT KAKENHI Grant Number 15H05888, 19H05110.
T.~Tenkanen was supported by the Simons foundation.
S.~Yokoyama was supported by MEXT KAKENHI Grant Numbers 15H05888 and 18H04356.
T.~Takahashi and S.~Yokoyama would like to thank JSPS and NRF under the Japan - Korea
Basic Scientific Cooperation Program for providing an opportunity for discussions.

\appendix

\section{More on the $\xi$-dependence of the behavior of the slow-roll parameter $\epsilon$}
\label{app:SRP_behavior}

As shown in Fig.~\ref{fig:PLI_ns_r}, for larger $\xi$, the tensor-to-scalar ratio asymptotically gets closer to the value in the minimally coupled case, that is, $r = 16 \epsilon = 8 \alpha^2$ for the PLI model.
In order to understand this feature better, here we investigate how the evolution of the slow-roll parameter $\epsilon$ changes depending on the non-minimal coupling parameter $\xi$.

Below we focus on the Palatini case because the differences between the metric and Palatini theories of gravity are only quite small, 
as can be seen in Fig.~\ref{fig:PLI_epsilon}, and the Palatini case is simpler to analyze.
Although here  we consider only the PLI model discussed in Sec. \ref{sec:nm_PLI}, the behavior in the case of the IMI model can be explained in the same way.

\begin{figure}
\begin{center}
\includegraphics[width=7.5cm]{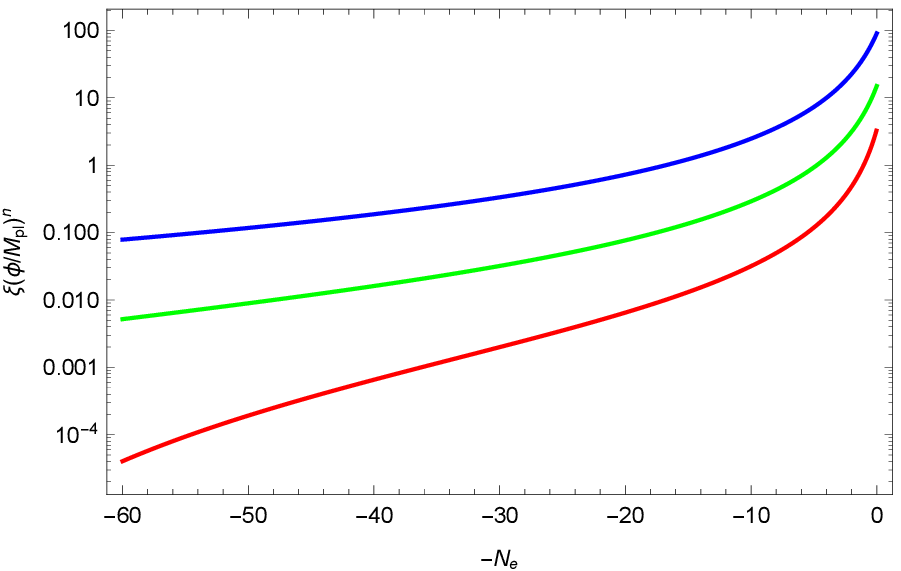}
\includegraphics[width=7.5cm]{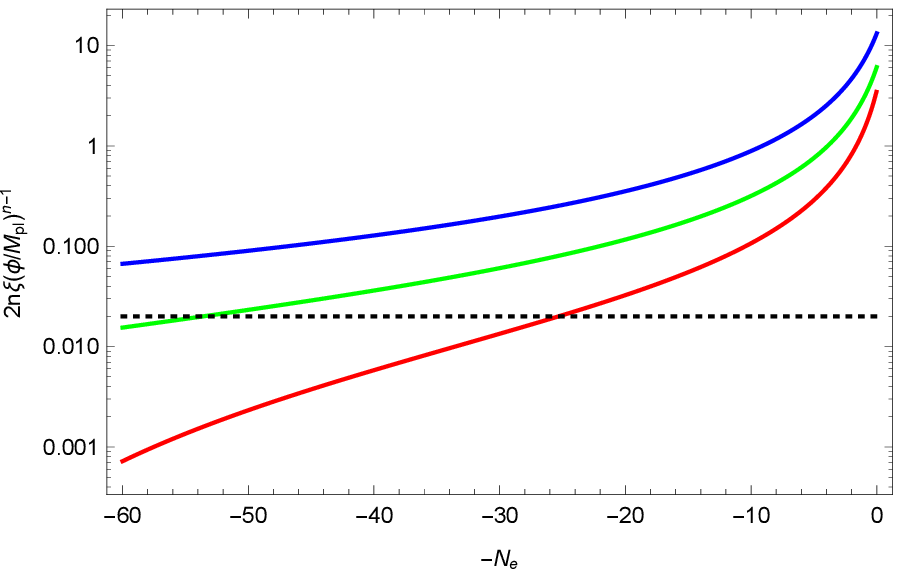}
\vspace{10mm}
\caption{Evolution of the two contributions: $\xi (\phi / M_{\rm pl})^n$ (left)
and $2n \xi (\phi/M_{\rm pl})^{n-1}$ (right)
as a function of $-N_{\rm e}$ in the PLI model. 
Here we take $\alpha = 0.02$. The blue curve is for $\xi = 10^{-5}$, whereas the green and red ones are for $\xi = 10^{-4}$ and $10^{-3}$, respectively. In the right panel, we show $\alpha = 0.02$ with a black dotted line for comparison.
}
\label{fig:PLI_mod}
\end{center}
\end{figure}

In the PLI model, the slow-roll parameter $\epsilon$, which is defined by
Eq.~(\ref{eq:SR_param}), is given by
\begin{equation}
\epsilon = \frac{1}{2} \left( 1 + \xi  \left(\frac{\phi}{M_{\rm pl}}\right)^n \right) \left[ \alpha + \frac{2n\xi(\phi/M_{\rm pl})^{n-1}}{1+\xi (\phi/M_{\rm pl})^n}   \right]^2~.
\end{equation}
In this expression,
one can see that the behavior of the slow-roll parameter during inflation
is characterized by two contributions from the non-minimal coupling,
$\xi (\phi / M_{\rm pl})^n$ and 
$2n \xi (\phi/M_{\rm Pl})^{n-1}$.
As can be seen in the left panel of Fig.~\ref{fig:PLI_mod}, the value of $\xi (\phi / M_{\rm pl})^n$  is 
smaller than unity around $N_{\rm e} \sim 50 - 60$ and thus the slow-roll parameter evaluated at $N_{\rm e} \sim 50 - 60$
is approximately given by
\begin{equation}
\epsilon \simeq \frac{1}{2}  \left( \alpha + 2n\xi\left(\frac{\phi}{M_{\rm pl}}\right)^{n-1}  \right)^2~.
\end{equation}
The value of 
$2n \xi (\phi/M_{\rm pl})^{n-1}$  is
larger than $\alpha = 0.02$ in the $\xi = 10^{-5}$ case, while in the $\xi = 10^{-3}$ case this  contribution is smaller than $\alpha$, as can be seen in the right panel of Fig.~\ref{fig:PLI_mod}.
Thus, for larger $\xi$, the tensor-to-scalar ratio, $r = 16 \epsilon$, approaches the value in the minimally coupled case, that is, $r = 8 \alpha^2$.

\begin{figure}
\begin{center}
\includegraphics[width=9cm]{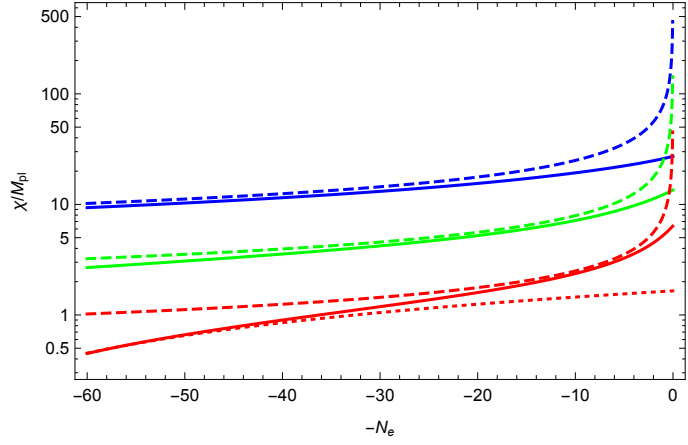}
\vspace{10mm}
\caption{Evolution of the $\chi$ field
as a function of $-N_{\rm e}$. 
Here we take $\alpha = 0.02$. 
The blue curve is for $\xi = 10^{-5}$, whereas the green and red ones are for $\xi = 10^{-4}$ and $10^{-3}$, respectively. 
 For comparison, the analytic solution for each case is shown as a dashed curve for $\alpha < 8 \xi \left(\chi/M_{\rm pl}\right)^{3}$
or a dotted curve for $\alpha > 8 \xi \left(\chi/M_{\rm pl}\right)^{3}$.
}
\label{fig:PLI_chi}
\end{center}
\end{figure}

Next, let us consider the reason why for larger $\xi$
we get a smaller contribution from the non-minimal coupling to the slow-roll parameter $\epsilon$ at $N_{\rm e} \sim 50 - 60$.
Assuming $\xi (\phi/M_{\rm pl})^n \ll 1$ which is basically guaranteed for $N_{\rm e} \sim 50 - 60$ as shown in Fig.~\ref{fig:PLI_mod}, we  have
\begin{equation}
    \frac{d\phi}{d\chi} \simeq 1~,~\frac{d}{dN}\left(\frac{\chi}{M_{\rm pl}} \right) \approx - M_{\rm pl}\frac{V_E'}{V_E} 
    \simeq \alpha + 8 \xi \left(\frac{\chi}{M_{\rm pl}}\right)^{3}~,
\end{equation}
where we have used $\phi \simeq \chi$ and $n=4$. 
For the $\alpha > 8 \xi \left(\chi/M_{\rm pl}\right)^{3}$ phase, 
the solution to the above equation is
\begin{equation}
\chi (N) \sim \alpha N M_{\rm pl}~.
\end{equation}
On the other hand, for the $\alpha < 8 \xi \left(\chi/M_{\rm pl}\right)^{3}$ phase we can obtain an approximate solution as
\begin{equation}
\chi (N_{\rm e}) \sim \frac{M_{\rm pl}}{(16 \xi N_{\rm e})^{1/2}}~.
\end{equation}
From this solution, we see that for fixed $N_{\rm e}$,
$\chi$ becomes smaller for larger $\xi$, and hence
the contribution from non-minimal coupling in the slow-roll parameter 
at $N_{\rm e} \sim 50 - 60$ gets smaller for larger $\xi$. In Fig.~\ref{fig:PLI_chi},
we compare the above analytic solutions to numerical ones,
and we find that the analytic solutions 
fit the numerical ones well.

\pagebreak
\section{Asymptotic form of the potential after inflation}
\label{app:IMI_eos}

In this appendix, we explain why the dynamics after inflation depends on the inflation model and also on the theory of gravity. 
Here we adopt units of $M_{\rm pl}=1$.

\begin{figure}[htbp]
\begin{center}
\includegraphics[width=7.5cm]{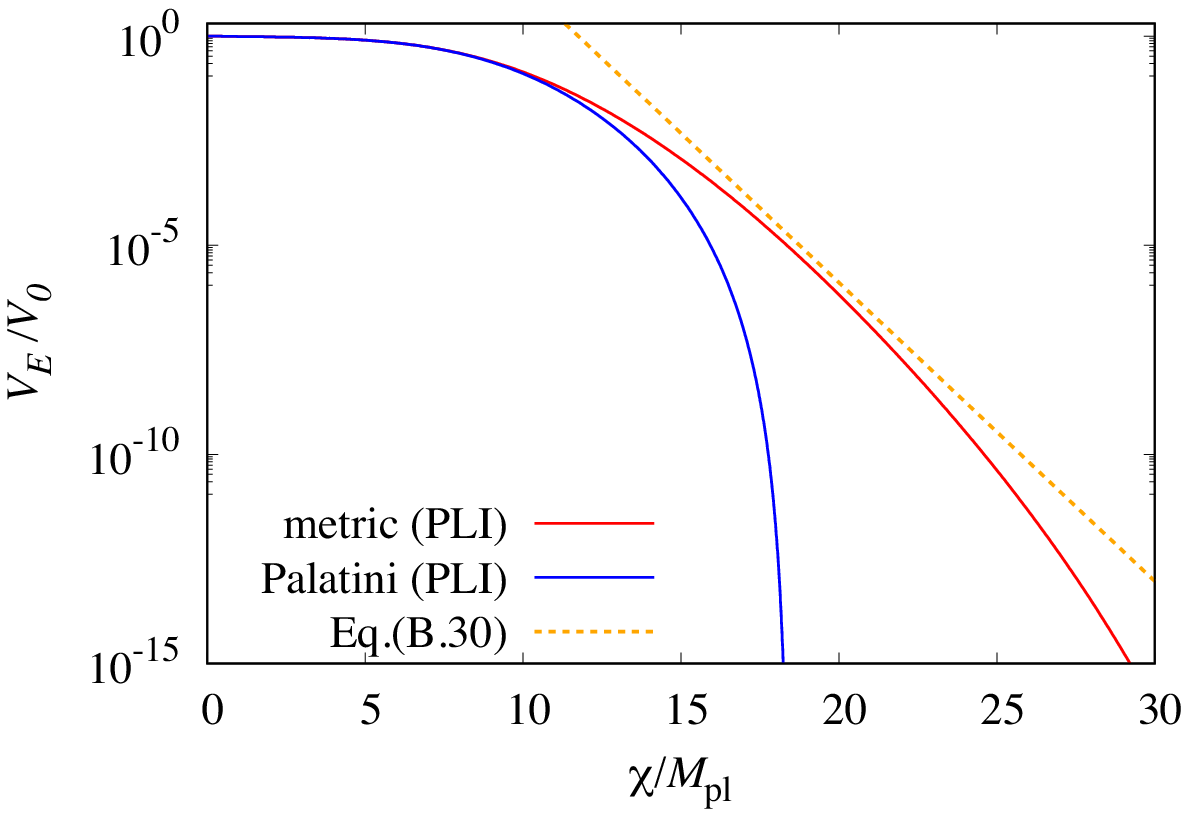}
\includegraphics[width=7.5cm]{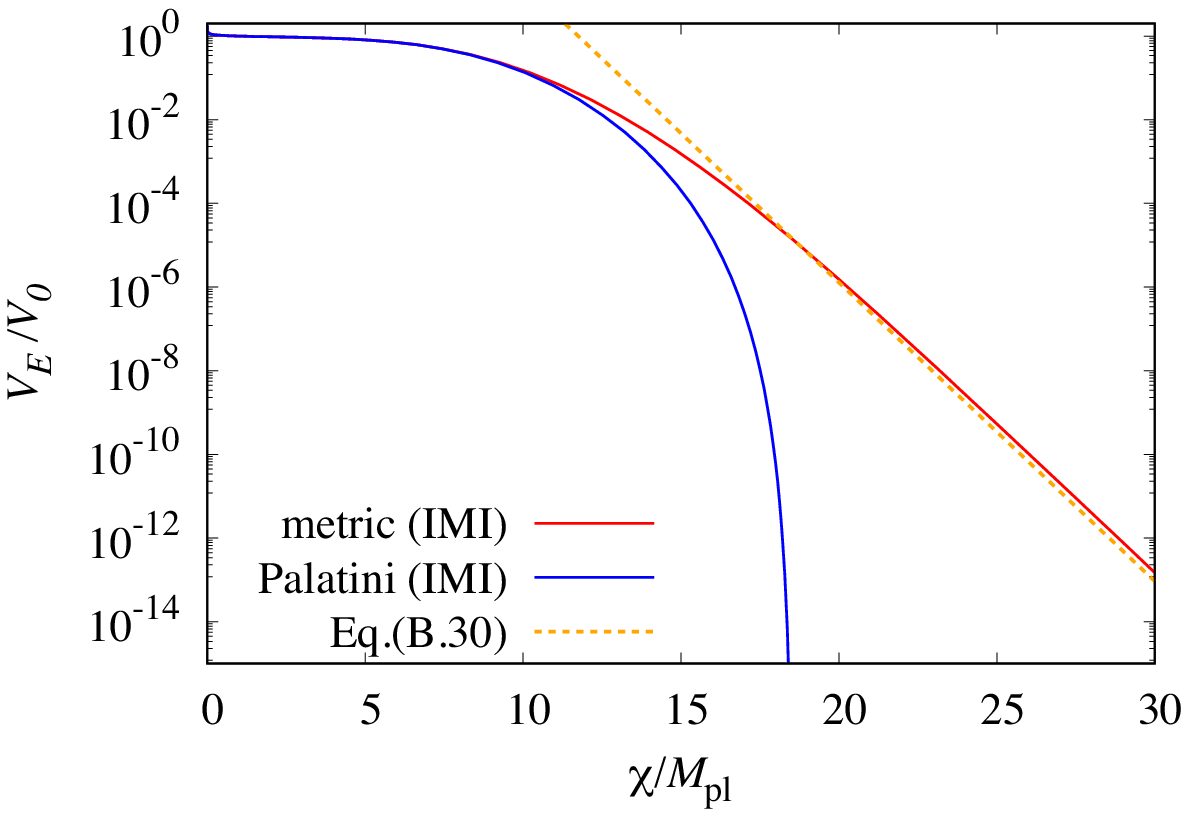}
\end{center}
\caption{Comparison of the Einstein frame potential between the metric and Palatini cases. In order to see the difference clearly at the large $\chi$ region, we 
show the results in a logarithmic scale. Left panel: $V_E/V_0$ in the PLI model with $\alpha = 0.02$ and $\xi =10^{-4}$.  Right panel: $V_E/V_0$ in the IMI model with $p = 0.05$ and $\xi =10^{-4}$.
By the orange dotted line, we show the approximate form given by Eq.~\eqref{eq:app}
with $\alpha = \sqrt{2/3} (p+8)/n \simeq 1.64$,
which corresponds to the asymptotic form of the potential in the IMI model. We also show the same line 
in the left panel for reference.
}
\label{fig:IMI_pot_3}
\end{figure}

Let us assume that after the end of inflation, $\phi \gg 1$ and the non-minimal coupling term is dominant.
This is the case in both the PLI and IMI models we have considered in this paper.
Then, we have
\begin{eqnarray}
\frac{d\phi}{d\chi} \simeq 
\begin{cases}
    \displaystyle\frac{\xi \phi^n}{\sqrt{\xi \phi^n}} = \xi^{1/2} \phi^{n/2}  & ({\rm Palatini})\, \\
    \displaystyle\frac{\xi \phi^n}{\sqrt{3/2 n^2 \xi^2 \phi^{2n-2}}} = \sqrt{\frac{2}{3}} \frac{1}{n} \phi  & ({\rm metric})\,.
  \end{cases}
\end{eqnarray}
For the Palatini case, with $n > 2$, we obtain a solution as
\begin{eqnarray}
\phi_0^{(2-n)/2} - \phi^{(2-n)/2} = \frac{n-2}{2} \xi^{1/2} (\chi - \chi_0),
\end{eqnarray}
and for the metric case, we have
\begin{eqnarray}
\phi = \phi_0 \exp \left[ \sqrt{\frac{2}{3}} \frac{1}{n} (\chi - \chi_0) \right]~.
\end{eqnarray}
Here $\phi_0$ and $\chi_0$ can be taken to be the field values at the end of inflation.
In the Palatini case, by taking $n=4$ as in the main text, we write the above solution as
\begin{eqnarray}
\phi = \frac{\xi^{-1/2}}{C-\chi},
\end{eqnarray}
where $C \equiv \xi^{-1/2} \phi_0^{-1} + \chi_0$.
Substituting this solution into the expression for the potential after inflation, for the PLI model we have
\begin{eqnarray}
V_E^{(P)} (\chi) \simeq  \frac{ V_0 \exp \left[ -\alpha \phi(\chi) \right]}{\xi^2 \phi(\chi)^8}
= V_0 \xi^2 (C - \chi)^8 \exp \left[ - \alpha \xi^{-1/2} (C - \chi)^{-1} \right]~,
\end{eqnarray}
and for the IMI model
\begin{eqnarray}
V_E^{(P)} (\chi) \simeq \frac{V_0 \phi(\chi)^{-p}}{\xi^2 \phi(\chi)^8}
= V_0 \xi^{(4+p)/2} (C-\chi)^{8+p}~.
\end{eqnarray}
The superscripts emphasize that these results apply for the Palatini theory. In both models, after the end of inflation $C-\chi$ quickly becomes small; that is, $\phi$ asymptotically diverges, and then the potential is quickly damped.
Thus, we expect to realize a kination phase.
In Fig.~\ref{fig:IMI_pot_3}, we show a numerical result for the potential in terms of $\chi$. In order to see the difference clearly at the large $\chi$ region, that is, after the end of inflation, we show the results in a logarithmic scale. In both panels, the blue curve is for the Palatini case. We see that in the Palatini case, the potential is indeed quickly damped soon after the end of inflation in both the PLI and IMI models.

In the same way, by substituting the solution in the metric case into the expression for the potential, we obtain for the PLI model with general $n$
\begin{eqnarray}
V_E^{(M)} (\chi) &\simeq& \frac{ V_0 \exp \left[ -\alpha \phi(\chi) \right]}{\xi^2 \phi(\chi)^{2n}} \\ \nonumber
&=& V_0 \xi^{-2} \phi_0^{-2n} \exp \left[ - \alpha  \phi_0 \exp \left[ \sqrt{\frac{2}{3}} \frac{1}{n} (\chi - \chi_0) \right] \right] \exp \left[ -2\sqrt{\frac{2}{3}} (\chi - \chi_0) \right]~.
\label{eq:V_PLI_metric}
\end{eqnarray}
As $\chi$ grows, this potential is also quickly damped and 
a kination phase can be realized.
As one can see from the above formula,
the potential for the PLI model in the metric case becomes steeper than a simple exponential potential.
On the other hand, in the metric IMI model we have
\begin{eqnarray}
V_E^{(M)} (\chi) \simeq \frac{V_0 \phi(\chi)^{-p}}{\xi^2 \phi(\chi)^{2n}}
= V_0 \xi^{-2} \phi_0^{-2n-p} \exp \left[ -\sqrt{\frac{2}{3}} \left(\frac{p}{n}+2 \right) (\chi - \chi_0) \right]~.
\label{eq:app}
\end{eqnarray}
This is just the potential of the minimal PLI model with $\alpha = \sqrt{2/3} (2 + p/n)$.
In the right panel of Fig.~\ref{fig:IMI_pot_3}, we assume $p=0.05$ and $\xi = 10^{-4}$.
The end of inflation in this figure corresponds to $\chi_0 \simeq 18 M_{\rm pl}$.
The orange dashed line is the approximate form given by Eq.~(\ref{eq:app}).  As a reference, the same line is also shown in the left panel for the PLI model. We 
see that the approximate form is  in agreement with the numerical result after the end of inflation.
Furthermore, in the minimal PLI model, we have an approximate relation between $\alpha$ and the equation of state $w_\chi$
as $\alpha^2 \simeq 3 (1+w_\chi)$, and from this equation, for our case with $p = 0.05$ and $n = 4$,
we have $\alpha \simeq 1.64$ and $w_\chi \simeq -0.1$.
This is again consistent with the numerically evaluated value of the equation-of-state parameter in the IMI case.

\clearpage 

\bibliography{non-minimal_inflation}

\begin{thebibliography}{76}
\expandafter\ifx\csname natexlab\endcsname\relax\def\natexlab#1{#1}\fi
\expandafter\ifx\csname bibnamefont\endcsname\relax
  \def\bibnamefont#1{#1}\fi
\expandafter\ifx\csname bibfnamefont\endcsname\relax
  \def\bibfnamefont#1{#1}\fi
\expandafter\ifx\csname citenamefont\endcsname\relax
  \def\citenamefont#1{#1}\fi
\expandafter\ifx\csname url\endcsname\relax
  \def\url#1{\texttt{#1}}\fi
\expandafter\ifx\csname urlprefix\endcsname\relax\def\urlprefix{URL }\fi
\providecommand{\bibinfo}[2]{#2}
\providecommand{\eprint}[2][]{\url{#2}}

\bibitem[{\citenamefont{Akrami et~al.}(2018)}]{Akrami:2018odb}
\bibinfo{author}{\bibfnamefont{Y.}~\bibnamefont{Akrami}} \bibnamefont{et~al.}
  (\bibinfo{collaboration}{Planck}) (\bibinfo{year}{2018}),
  \eprint{1807.06211}.

\bibitem[{\citenamefont{Martin et~al.}(2014)\citenamefont{Martin, Ringeval, and
  Vennin}}]{Martin:2013tda}
\bibinfo{author}{\bibfnamefont{J.}~\bibnamefont{Martin}},
  \bibinfo{author}{\bibfnamefont{C.}~\bibnamefont{Ringeval}}, \bibnamefont{and}
  \bibinfo{author}{\bibfnamefont{V.}~\bibnamefont{Vennin}},
  \bibinfo{journal}{Phys. Dark Univ.} \textbf{\bibinfo{volume}{5-6}},
  \bibinfo{pages}{75} (\bibinfo{year}{2014}), \eprint{1303.3787}.

\bibitem[{\citenamefont{Enqvist and Sloth}(2002)}]{Enqvist:2001zp}
\bibinfo{author}{\bibfnamefont{K.}~\bibnamefont{Enqvist}} \bibnamefont{and}
  \bibinfo{author}{\bibfnamefont{M.~S.} \bibnamefont{Sloth}},
  \bibinfo{journal}{Nucl. Phys.} \textbf{\bibinfo{volume}{B626}},
  \bibinfo{pages}{395} (\bibinfo{year}{2002}), \eprint{hep-ph/0109214}.

\bibitem[{\citenamefont{Lyth and Wands}(2002)}]{Lyth:2001nq}
\bibinfo{author}{\bibfnamefont{D.~H.} \bibnamefont{Lyth}} \bibnamefont{and}
  \bibinfo{author}{\bibfnamefont{D.}~\bibnamefont{Wands}},
  \bibinfo{journal}{Phys. Lett.} \textbf{\bibinfo{volume}{B524}},
  \bibinfo{pages}{5} (\bibinfo{year}{2002}), \eprint{hep-ph/0110002}.

\bibitem[{\citenamefont{Moroi and Takahashi}(2001)}]{Moroi:2001ct}
\bibinfo{author}{\bibfnamefont{T.}~\bibnamefont{Moroi}} \bibnamefont{and}
  \bibinfo{author}{\bibfnamefont{T.}~\bibnamefont{Takahashi}},
  \bibinfo{journal}{Phys. Lett.} \textbf{\bibinfo{volume}{B522}},
  \bibinfo{pages}{215} (\bibinfo{year}{2001}), \bibinfo{note}{[Erratum: Phys.
  Lett.B539,303(2002)]}, \eprint{hep-ph/0110096}.

\bibitem[{\citenamefont{Dvali et~al.}(2004)\citenamefont{Dvali, Gruzinov, and
  Zaldarriaga}}]{Dvali:2003em}
\bibinfo{author}{\bibfnamefont{G.}~\bibnamefont{Dvali}},
  \bibinfo{author}{\bibfnamefont{A.}~\bibnamefont{Gruzinov}}, \bibnamefont{and}
  \bibinfo{author}{\bibfnamefont{M.}~\bibnamefont{Zaldarriaga}},
  \bibinfo{journal}{Phys. Rev.} \textbf{\bibinfo{volume}{D69}},
  \bibinfo{pages}{023505} (\bibinfo{year}{2004}), \eprint{astro-ph/0303591}.

\bibitem[{\citenamefont{Kofman}(2003)}]{Kofman:2003nx}
\bibinfo{author}{\bibfnamefont{L.}~\bibnamefont{Kofman}}
  (\bibinfo{year}{2003}), \eprint{astro-ph/0303614}.

\bibitem[{\citenamefont{Langlois and Vernizzi}(2004)}]{Langlois:2004nn}
\bibinfo{author}{\bibfnamefont{D.}~\bibnamefont{Langlois}} \bibnamefont{and}
  \bibinfo{author}{\bibfnamefont{F.}~\bibnamefont{Vernizzi}},
  \bibinfo{journal}{Phys. Rev.} \textbf{\bibinfo{volume}{D70}},
  \bibinfo{pages}{063522} (\bibinfo{year}{2004}), \eprint{astro-ph/0403258}.

\bibitem[{\citenamefont{Moroi et~al.}(2005)\citenamefont{Moroi, Takahashi, and
  Toyoda}}]{Moroi:2005kz}
\bibinfo{author}{\bibfnamefont{T.}~\bibnamefont{Moroi}},
  \bibinfo{author}{\bibfnamefont{T.}~\bibnamefont{Takahashi}},
  \bibnamefont{and} \bibinfo{author}{\bibfnamefont{Y.}~\bibnamefont{Toyoda}},
  \bibinfo{journal}{Phys. Rev.} \textbf{\bibinfo{volume}{D72}},
  \bibinfo{pages}{023502} (\bibinfo{year}{2005}), \eprint{hep-ph/0501007}.

\bibitem[{\citenamefont{Moroi and Takahashi}(2005)}]{Moroi:2005np}
\bibinfo{author}{\bibfnamefont{T.}~\bibnamefont{Moroi}} \bibnamefont{and}
  \bibinfo{author}{\bibfnamefont{T.}~\bibnamefont{Takahashi}},
  \bibinfo{journal}{Phys. Rev.} \textbf{\bibinfo{volume}{D72}},
  \bibinfo{pages}{023505} (\bibinfo{year}{2005}), \eprint{astro-ph/0505339}.

\bibitem[{\citenamefont{Ichikawa et~al.}(2008)\citenamefont{Ichikawa, Suyama,
  Takahashi, and Yamaguchi}}]{Ichikawa:2008iq}
\bibinfo{author}{\bibfnamefont{K.}~\bibnamefont{Ichikawa}},
  \bibinfo{author}{\bibfnamefont{T.}~\bibnamefont{Suyama}},
  \bibinfo{author}{\bibfnamefont{T.}~\bibnamefont{Takahashi}},
  \bibnamefont{and}
  \bibinfo{author}{\bibfnamefont{M.}~\bibnamefont{Yamaguchi}},
  \bibinfo{journal}{Phys. Rev.} \textbf{\bibinfo{volume}{D78}},
  \bibinfo{pages}{023513} (\bibinfo{year}{2008}), \eprint{0802.4138}.

\bibitem[{\citenamefont{Enqvist and Takahashi}(2013)}]{Enqvist:2013paa}
\bibinfo{author}{\bibfnamefont{K.}~\bibnamefont{Enqvist}} \bibnamefont{and}
  \bibinfo{author}{\bibfnamefont{T.}~\bibnamefont{Takahashi}},
  \bibinfo{journal}{JCAP} \textbf{\bibinfo{volume}{1310}}, \bibinfo{pages}{034}
  (\bibinfo{year}{2013}), \eprint{1306.5958}.

\bibitem[{\citenamefont{Vennin et~al.}(2015)\citenamefont{Vennin, Koyama, and
  Wands}}]{Vennin:2015vfa}
\bibinfo{author}{\bibfnamefont{V.}~\bibnamefont{Vennin}},
  \bibinfo{author}{\bibfnamefont{K.}~\bibnamefont{Koyama}}, \bibnamefont{and}
  \bibinfo{author}{\bibfnamefont{D.}~\bibnamefont{Wands}},
  \bibinfo{journal}{JCAP} \textbf{\bibinfo{volume}{1511}}, \bibinfo{pages}{008}
  (\bibinfo{year}{2015}), \eprint{1507.07575}.

\bibitem[{\citenamefont{Haba et~al.}(2018)\citenamefont{Haba, Takahashi, and
  Yamada}}]{Haba:2017fbi}
\bibinfo{author}{\bibfnamefont{N.}~\bibnamefont{Haba}},
  \bibinfo{author}{\bibfnamefont{T.}~\bibnamefont{Takahashi}},
  \bibnamefont{and} \bibinfo{author}{\bibfnamefont{T.}~\bibnamefont{Yamada}},
  \bibinfo{journal}{JCAP} \textbf{\bibinfo{volume}{1806}}, \bibinfo{pages}{011}
  (\bibinfo{year}{2018}), \eprint{1712.03684}.

\bibitem[{\citenamefont{Bezrukov and Shaposhnikov}(2008)}]{Bezrukov:2007ep}
\bibinfo{author}{\bibfnamefont{F.~L.} \bibnamefont{Bezrukov}} \bibnamefont{and}
  \bibinfo{author}{\bibfnamefont{M.}~\bibnamefont{Shaposhnikov}},
  \bibinfo{journal}{Phys. Lett.} \textbf{\bibinfo{volume}{B659}},
  \bibinfo{pages}{703} (\bibinfo{year}{2008}), \eprint{0710.3755}.

\bibitem[{\citenamefont{Spokoiny}(1984)}]{Spokoiny:1984bd}
\bibinfo{author}{\bibfnamefont{B.~L.} \bibnamefont{Spokoiny}},
  \bibinfo{journal}{Phys. Lett.} \textbf{\bibinfo{volume}{147B}},
  \bibinfo{pages}{39} (\bibinfo{year}{1984}).

\bibitem[{\citenamefont{Futamase and Maeda}(1989)}]{Futamase:1987ua}
\bibinfo{author}{\bibfnamefont{T.}~\bibnamefont{Futamase}} \bibnamefont{and}
  \bibinfo{author}{\bibfnamefont{K.-i.} \bibnamefont{Maeda}},
  \bibinfo{journal}{Phys. Rev.} \textbf{\bibinfo{volume}{D39}},
  \bibinfo{pages}{399} (\bibinfo{year}{1989}).

\bibitem[{\citenamefont{Salopek et~al.}(1989)\citenamefont{Salopek, Bond, and
  Bardeen}}]{Salopek:1988qh}
\bibinfo{author}{\bibfnamefont{D.~S.} \bibnamefont{Salopek}},
  \bibinfo{author}{\bibfnamefont{J.~R.} \bibnamefont{Bond}}, \bibnamefont{and}
  \bibinfo{author}{\bibfnamefont{J.~M.} \bibnamefont{Bardeen}},
  \bibinfo{journal}{Phys. Rev.} \textbf{\bibinfo{volume}{D40}},
  \bibinfo{pages}{1753} (\bibinfo{year}{1989}).

\bibitem[{\citenamefont{Fakir and Unruh}(1990)}]{Fakir:1990eg}
\bibinfo{author}{\bibfnamefont{R.}~\bibnamefont{Fakir}} \bibnamefont{and}
  \bibinfo{author}{\bibfnamefont{W.~G.} \bibnamefont{Unruh}},
  \bibinfo{journal}{Phys. Rev.} \textbf{\bibinfo{volume}{D41}},
  \bibinfo{pages}{1783} (\bibinfo{year}{1990}).

\bibitem[{\citenamefont{Amendola et~al.}(1990)\citenamefont{Amendola, Litterio,
  and Occhionero}}]{Amendola:1990nn}
\bibinfo{author}{\bibfnamefont{L.}~\bibnamefont{Amendola}},
  \bibinfo{author}{\bibfnamefont{M.}~\bibnamefont{Litterio}}, \bibnamefont{and}
  \bibinfo{author}{\bibfnamefont{F.}~\bibnamefont{Occhionero}},
  \bibinfo{journal}{Int. J. Mod. Phys.} \textbf{\bibinfo{volume}{A5}},
  \bibinfo{pages}{3861} (\bibinfo{year}{1990}).

\bibitem[{\citenamefont{Kaiser}(1995)}]{Kaiser:1994vs}
\bibinfo{author}{\bibfnamefont{D.~I.} \bibnamefont{Kaiser}},
  \bibinfo{journal}{Phys. Rev.} \textbf{\bibinfo{volume}{D52}},
  \bibinfo{pages}{4295} (\bibinfo{year}{1995}), \eprint{astro-ph/9408044}.

\bibitem[{\citenamefont{Cervantes-Cota and
  Dehnen}(1995)}]{CervantesCota:1995tz}
\bibinfo{author}{\bibfnamefont{J.~L.} \bibnamefont{Cervantes-Cota}}
  \bibnamefont{and} \bibinfo{author}{\bibfnamefont{H.}~\bibnamefont{Dehnen}},
  \bibinfo{journal}{Nucl. Phys.} \textbf{\bibinfo{volume}{B442}},
  \bibinfo{pages}{391} (\bibinfo{year}{1995}), \eprint{astro-ph/9505069}.

\bibitem[{\citenamefont{Komatsu and Futamase}(1999)}]{Komatsu:1999mt}
\bibinfo{author}{\bibfnamefont{E.}~\bibnamefont{Komatsu}} \bibnamefont{and}
  \bibinfo{author}{\bibfnamefont{T.}~\bibnamefont{Futamase}},
  \bibinfo{journal}{Phys. Rev.} \textbf{\bibinfo{volume}{D59}},
  \bibinfo{pages}{064029} (\bibinfo{year}{1999}), \eprint{astro-ph/9901127}.

\bibitem[{\citenamefont{Rubio}(2019)}]{Rubio:2018ogq}
\bibinfo{author}{\bibfnamefont{J.}~\bibnamefont{Rubio}},
  \bibinfo{journal}{Front. Astron. Space Sci.} \textbf{\bibinfo{volume}{5}},
  \bibinfo{pages}{50} (\bibinfo{year}{2019}), \eprint{1807.02376}.

\bibitem[{\citenamefont{Tenkanen}(2020)}]{Tenkanen:2020dge}
\bibinfo{author}{\bibfnamefont{T.}~\bibnamefont{Tenkanen}},
  \bibinfo{journal}{Gen. Rel. Grav.} \textbf{\bibinfo{volume}{52}},
  \bibinfo{pages}{33} (\bibinfo{year}{2020}), \eprint{2001.10135}.

\bibitem[{\citenamefont{Germani and Kehagias}(2010)}]{Germani:2010gm}
\bibinfo{author}{\bibfnamefont{C.}~\bibnamefont{Germani}} \bibnamefont{and}
  \bibinfo{author}{\bibfnamefont{A.}~\bibnamefont{Kehagias}},
  \bibinfo{journal}{Phys. Rev. Lett.} \textbf{\bibinfo{volume}{105}},
  \bibinfo{pages}{011302} (\bibinfo{year}{2010}), \eprint{1003.2635}.

\bibitem[{\citenamefont{Granda}(2011)}]{Granda:2011zk}
\bibinfo{author}{\bibfnamefont{L.}~\bibnamefont{Granda}},
  \bibinfo{journal}{JCAP} \textbf{\bibinfo{volume}{04}}, \bibinfo{pages}{016}
  (\bibinfo{year}{2011}), \eprint{1104.2253}.

\bibitem[{\citenamefont{Kamada et~al.}(2012)\citenamefont{Kamada, Kobayashi,
  Takahashi, Yamaguchi, and Yokoyama}}]{Kamada:2012se}
\bibinfo{author}{\bibfnamefont{K.}~\bibnamefont{Kamada}},
  \bibinfo{author}{\bibfnamefont{T.}~\bibnamefont{Kobayashi}},
  \bibinfo{author}{\bibfnamefont{T.}~\bibnamefont{Takahashi}},
  \bibinfo{author}{\bibfnamefont{M.}~\bibnamefont{Yamaguchi}},
  \bibnamefont{and} \bibinfo{author}{\bibfnamefont{J.}~\bibnamefont{Yokoyama}},
  \bibinfo{journal}{Phys. Rev. D} \textbf{\bibinfo{volume}{86}},
  \bibinfo{pages}{023504} (\bibinfo{year}{2012}), \eprint{1203.4059}.

\bibitem[{\citenamefont{Lerner and McDonald}(2011)}]{Lerner:2011ge}
\bibinfo{author}{\bibfnamefont{R.~N.} \bibnamefont{Lerner}} \bibnamefont{and}
  \bibinfo{author}{\bibfnamefont{J.}~\bibnamefont{McDonald}},
  \bibinfo{journal}{Phys. Rev.} \textbf{\bibinfo{volume}{D83}},
  \bibinfo{pages}{123522} (\bibinfo{year}{2011}), \eprint{1104.2468}.

\bibitem[{\citenamefont{Bezrukov and Gorbunov}(2012)}]{Bezrukov:2011gp}
\bibinfo{author}{\bibfnamefont{F.~L.} \bibnamefont{Bezrukov}} \bibnamefont{and}
  \bibinfo{author}{\bibfnamefont{D.~S.} \bibnamefont{Gorbunov}},
  \bibinfo{journal}{Phys. Lett.} \textbf{\bibinfo{volume}{B713}},
  \bibinfo{pages}{365} (\bibinfo{year}{2012}), \eprint{1111.4397}.

\bibitem[{\citenamefont{Kaiser and Sfakianakis}(2014)}]{Kaiser:2013sna}
\bibinfo{author}{\bibfnamefont{D.~I.} \bibnamefont{Kaiser}} \bibnamefont{and}
  \bibinfo{author}{\bibfnamefont{E.~I.} \bibnamefont{Sfakianakis}},
  \bibinfo{journal}{Phys. Rev. Lett.} \textbf{\bibinfo{volume}{112}},
  \bibinfo{pages}{011302} (\bibinfo{year}{2014}), \eprint{1304.0363}.

\bibitem[{\citenamefont{Kallosh et~al.}(2014)\citenamefont{Kallosh, Linde, and
  Roest}}]{Kallosh:2013tua}
\bibinfo{author}{\bibfnamefont{R.}~\bibnamefont{Kallosh}},
  \bibinfo{author}{\bibfnamefont{A.}~\bibnamefont{Linde}}, \bibnamefont{and}
  \bibinfo{author}{\bibfnamefont{D.}~\bibnamefont{Roest}},
  \bibinfo{journal}{Phys. Rev. Lett.} \textbf{\bibinfo{volume}{112}},
  \bibinfo{pages}{011303} (\bibinfo{year}{2014}), \eprint{1310.3950}.

\bibitem[{\citenamefont{Gong et~al.}(2015)\citenamefont{Gong, Pi, and
  Leung}}]{Gong:2015qha}
\bibinfo{author}{\bibfnamefont{J.-O.} \bibnamefont{Gong}},
  \bibinfo{author}{\bibfnamefont{S.}~\bibnamefont{Pi}}, \bibnamefont{and}
  \bibinfo{author}{\bibfnamefont{G.}~\bibnamefont{Leung}},
  \bibinfo{journal}{JCAP} \textbf{\bibinfo{volume}{05}}, \bibinfo{pages}{027}
  (\bibinfo{year}{2015}), \eprint{1501.03604}.

\bibitem[{\citenamefont{Takahashi and Tenkanen}(2019)}]{Takahashi:2018brt}
\bibinfo{author}{\bibfnamefont{T.}~\bibnamefont{Takahashi}} \bibnamefont{and}
  \bibinfo{author}{\bibfnamefont{T.}~\bibnamefont{Tenkanen}},
  \bibinfo{journal}{JCAP} \textbf{\bibinfo{volume}{1904}}, \bibinfo{pages}{035}
  (\bibinfo{year}{2019}), \eprint{1812.08492}.

\bibitem[{\citenamefont{Abbott and Wise}(1984)}]{Abbott:1984fp}
\bibinfo{author}{\bibfnamefont{L.~F.} \bibnamefont{Abbott}} \bibnamefont{and}
  \bibinfo{author}{\bibfnamefont{M.~B.} \bibnamefont{Wise}},
  \bibinfo{journal}{Nucl. Phys.} \textbf{\bibinfo{volume}{B244}},
  \bibinfo{pages}{541} (\bibinfo{year}{1984}).

\bibitem[{\citenamefont{Ratra and Peebles}(1988)}]{Ratra:1987rm}
\bibinfo{author}{\bibfnamefont{B.}~\bibnamefont{Ratra}} \bibnamefont{and}
  \bibinfo{author}{\bibfnamefont{P.~J.~E.} \bibnamefont{Peebles}},
  \bibinfo{journal}{Phys. Rev.} \textbf{\bibinfo{volume}{D37}},
  \bibinfo{pages}{3406} (\bibinfo{year}{1988}).

\bibitem[{\citenamefont{Tashiro et~al.}(2004)\citenamefont{Tashiro, Chiba, and
  Sasaki}}]{Tashiro:2003qp}
\bibinfo{author}{\bibfnamefont{H.}~\bibnamefont{Tashiro}},
  \bibinfo{author}{\bibfnamefont{T.}~\bibnamefont{Chiba}}, \bibnamefont{and}
  \bibinfo{author}{\bibfnamefont{M.}~\bibnamefont{Sasaki}},
  \bibinfo{journal}{Class. Quant. Grav.} \textbf{\bibinfo{volume}{21}},
  \bibinfo{pages}{1761} (\bibinfo{year}{2004}), \eprint{gr-qc/0307068}.

\bibitem[{\citenamefont{Ford}(1987)}]{Ford}
\bibinfo{author}{\bibfnamefont{L.~H.} \bibnamefont{Ford}},
  \bibinfo{journal}{Phys. Rev. D} \textbf{\bibinfo{volume}{35}},
  \bibinfo{pages}{2955} (\bibinfo{year}{1987}).

\bibitem[{\citenamefont{Starobinsky and Yokoyama}(1994)}]{Starobinsky:1994bd}
\bibinfo{author}{\bibfnamefont{A.~A.} \bibnamefont{Starobinsky}}
  \bibnamefont{and} \bibinfo{author}{\bibfnamefont{J.}~\bibnamefont{Yokoyama}},
  \bibinfo{journal}{Phys. Rev.} \textbf{\bibinfo{volume}{D50}},
  \bibinfo{pages}{6357} (\bibinfo{year}{1994}), \eprint{astro-ph/9407016}.

\bibitem[{\citenamefont{Bauer and Demir}(2008)}]{Bauer:2008zj}
\bibinfo{author}{\bibfnamefont{F.}~\bibnamefont{Bauer}} \bibnamefont{and}
  \bibinfo{author}{\bibfnamefont{D.~A.} \bibnamefont{Demir}},
  \bibinfo{journal}{Phys. Lett.} \textbf{\bibinfo{volume}{B665}},
  \bibinfo{pages}{222} (\bibinfo{year}{2008}), \eprint{0803.2664}.

\bibitem[{\citenamefont{Rasanen}(2018)}]{Rasanen:2018ihz}
\bibinfo{author}{\bibfnamefont{S.}~\bibnamefont{Rasanen}},
  \bibinfo{journal}{The Open Journal of Astrophysics}  (\bibinfo{year}{2018}),
  \eprint{1811.09514}.

\bibitem[{\citenamefont{Aoki and Mukohyama}(2020)}]{Aoki:2020zqm}
\bibinfo{author}{\bibfnamefont{K.}~\bibnamefont{Aoki}} \bibnamefont{and}
  \bibinfo{author}{\bibfnamefont{S.}~\bibnamefont{Mukohyama}}
  (\bibinfo{year}{2020}), \eprint{2003.00664}.

\bibitem[{\citenamefont{Garcia-Bellido
  et~al.}(2009)\citenamefont{Garcia-Bellido, Figueroa, and
  Rubio}}]{GarciaBellido:2008ab}
\bibinfo{author}{\bibfnamefont{J.}~\bibnamefont{Garcia-Bellido}},
  \bibinfo{author}{\bibfnamefont{D.~G.} \bibnamefont{Figueroa}},
  \bibnamefont{and} \bibinfo{author}{\bibfnamefont{J.}~\bibnamefont{Rubio}},
  \bibinfo{journal}{Phys. Rev.} \textbf{\bibinfo{volume}{D79}},
  \bibinfo{pages}{063531} (\bibinfo{year}{2009}), \eprint{0812.4624}.

\bibitem[{\citenamefont{Rasanen and Wahlman}(2017)}]{Rasanen:2017ivk}
\bibinfo{author}{\bibfnamefont{S.}~\bibnamefont{Rasanen}} \bibnamefont{and}
  \bibinfo{author}{\bibfnamefont{P.}~\bibnamefont{Wahlman}},
  \bibinfo{journal}{JCAP} \textbf{\bibinfo{volume}{1711}}, \bibinfo{pages}{047}
  (\bibinfo{year}{2017}), \eprint{1709.07853}.

\bibitem[{\citenamefont{Jarv et~al.}(2018)\citenamefont{Jarv, Racioppi, and
  Tenkanen}}]{Jarv:2017azx}
\bibinfo{author}{\bibfnamefont{L.}~\bibnamefont{Jarv}},
  \bibinfo{author}{\bibfnamefont{A.}~\bibnamefont{Racioppi}}, \bibnamefont{and}
  \bibinfo{author}{\bibfnamefont{T.}~\bibnamefont{Tenkanen}},
  \bibinfo{journal}{Phys. Rev.} \textbf{\bibinfo{volume}{D97}},
  \bibinfo{pages}{083513} (\bibinfo{year}{2018}), \eprint{1712.08471}.

\bibitem[{\citenamefont{Ade et~al.}(2018{\natexlab{a}})}]{Ade:2018gkx}
\bibinfo{author}{\bibfnamefont{P.~A.~R.} \bibnamefont{Ade}}
  \bibnamefont{et~al.} (\bibinfo{collaboration}{BICEP2, Keck Array}),
  \bibinfo{journal}{Submitted to: Phys. Rev. Lett.}
  (\bibinfo{year}{2018}{\natexlab{a}}), \eprint{1810.05216}.

\bibitem[{\citenamefont{Becker et~al.}(2005)\citenamefont{Becker, Becker, and
  Krause}}]{Becker:2005sg}
\bibinfo{author}{\bibfnamefont{K.}~\bibnamefont{Becker}},
  \bibinfo{author}{\bibfnamefont{M.}~\bibnamefont{Becker}}, \bibnamefont{and}
  \bibinfo{author}{\bibfnamefont{A.}~\bibnamefont{Krause}},
  \bibinfo{journal}{Nucl. Phys.} \textbf{\bibinfo{volume}{B715}},
  \bibinfo{pages}{349} (\bibinfo{year}{2005}), \eprint{hep-th/0501130}.

\bibitem[{\citenamefont{Peebles and Ratra}(1988)}]{Peebles:1987ek}
\bibinfo{author}{\bibfnamefont{P.~J.~E.} \bibnamefont{Peebles}}
  \bibnamefont{and} \bibinfo{author}{\bibfnamefont{B.}~\bibnamefont{Ratra}},
  \bibinfo{journal}{Astrophys. J.} \textbf{\bibinfo{volume}{325}},
  \bibinfo{pages}{L17} (\bibinfo{year}{1988}).

\bibitem[{\citenamefont{Barrow and Liddle}(1993)}]{Barrow:1993zq}
\bibinfo{author}{\bibfnamefont{J.~D.} \bibnamefont{Barrow}} \bibnamefont{and}
  \bibinfo{author}{\bibfnamefont{A.~R.} \bibnamefont{Liddle}},
  \bibinfo{journal}{Phys. Rev.} \textbf{\bibinfo{volume}{D47}},
  \bibinfo{pages}{R5219} (\bibinfo{year}{1993}), \eprint{astro-ph/9303011}.

\bibitem[{\citenamefont{Feinstein}(2002)}]{Feinstein:2002aj}
\bibinfo{author}{\bibfnamefont{A.}~\bibnamefont{Feinstein}},
  \bibinfo{journal}{Phys. Rev.} \textbf{\bibinfo{volume}{D66}},
  \bibinfo{pages}{063511} (\bibinfo{year}{2002}), \eprint{hep-th/0204140}.

\bibitem[{\citenamefont{Sami}(2003)}]{Sami:2002zy}
\bibinfo{author}{\bibfnamefont{M.}~\bibnamefont{Sami}}, \bibinfo{journal}{Mod.
  Phys. Lett.} \textbf{\bibinfo{volume}{A18}}, \bibinfo{pages}{691}
  (\bibinfo{year}{2003}), \eprint{hep-th/0205146}.

\bibitem[{\citenamefont{Kinney and Riotto}(1999)}]{Kinney:1997hm}
\bibinfo{author}{\bibfnamefont{W.~H.} \bibnamefont{Kinney}} \bibnamefont{and}
  \bibinfo{author}{\bibfnamefont{A.}~\bibnamefont{Riotto}},
  \bibinfo{journal}{Astropart. Phys.} \textbf{\bibinfo{volume}{10}},
  \bibinfo{pages}{387} (\bibinfo{year}{1999}), \eprint{hep-ph/9704388}.

\bibitem[{\citenamefont{Kinney and Riotto}(1998)}]{Kinney:1998dv}
\bibinfo{author}{\bibfnamefont{W.~H.} \bibnamefont{Kinney}} \bibnamefont{and}
  \bibinfo{author}{\bibfnamefont{A.}~\bibnamefont{Riotto}},
  \bibinfo{journal}{Phys. Lett.} \textbf{\bibinfo{volume}{B435}},
  \bibinfo{pages}{272} (\bibinfo{year}{1998}), \eprint{hep-ph/9802443}.

\bibitem[{\citenamefont{Frewin and Lidsey}(1993)}]{Frewin:1993aj}
\bibinfo{author}{\bibfnamefont{R.~A.} \bibnamefont{Frewin}} \bibnamefont{and}
  \bibinfo{author}{\bibfnamefont{J.~E.} \bibnamefont{Lidsey}},
  \bibinfo{journal}{Int. J. Mod. Phys.} \textbf{\bibinfo{volume}{D2}},
  \bibinfo{pages}{323} (\bibinfo{year}{1993}), \eprint{astro-ph/9312035}.

\bibitem[{\citenamefont{Kaganovich}(2001)}]{Kaganovich:2000fc}
\bibinfo{author}{\bibfnamefont{A.~B.} \bibnamefont{Kaganovich}},
  \bibinfo{journal}{Phys. Rev.} \textbf{\bibinfo{volume}{D63}},
  \bibinfo{pages}{025022} (\bibinfo{year}{2001}), \eprint{hep-th/0007144}.

\bibitem[{\citenamefont{Liddle and Leach}(2003)}]{Liddle:2003as}
\bibinfo{author}{\bibfnamefont{A.~R.} \bibnamefont{Liddle}} \bibnamefont{and}
  \bibinfo{author}{\bibfnamefont{S.~M.} \bibnamefont{Leach}},
  \bibinfo{journal}{Phys. Rev.} \textbf{\bibinfo{volume}{D68}},
  \bibinfo{pages}{103503} (\bibinfo{year}{2003}), \eprint{astro-ph/0305263}.

\bibitem[{\citenamefont{Tenkanen}(2017)}]{Tenkanen:2017jih}
\bibinfo{author}{\bibfnamefont{T.}~\bibnamefont{Tenkanen}},
  \bibinfo{journal}{JCAP} \textbf{\bibinfo{volume}{1712}}, \bibinfo{pages}{001}
  (\bibinfo{year}{2017}), \eprint{1710.02758}.

\bibitem[{\citenamefont{Racioppi}(2017)}]{Racioppi:2017spw}
\bibinfo{author}{\bibfnamefont{A.}~\bibnamefont{Racioppi}},
  \bibinfo{journal}{JCAP} \textbf{\bibinfo{volume}{1712}}, \bibinfo{pages}{041}
  (\bibinfo{year}{2017}), \eprint{1710.04853}.

\bibitem[{\citenamefont{Wu et~al.}(2016)}]{Wu:2016hul}
\bibinfo{author}{\bibfnamefont{W.~L.~K.} \bibnamefont{Wu}}
  \bibnamefont{et~al.}, \bibinfo{journal}{J. Low. Temp. Phys.}
  \textbf{\bibinfo{volume}{184}}, \bibinfo{pages}{765} (\bibinfo{year}{2016}),
  \eprint{1601.00125}.

\bibitem[{\citenamefont{Matsumura et~al.}(2013)}]{Matsumura:2013aja}
\bibinfo{author}{\bibfnamefont{T.}~\bibnamefont{Matsumura}}
  \bibnamefont{et~al.} (\bibinfo{year}{2013}), \bibinfo{note}{[J. Low. Temp.
  Phys.176,733(2014)]}, \eprint{1311.2847}.

\bibitem[{\citenamefont{Ade et~al.}(2018{\natexlab{b}})}]{Simons_Observatory}
\bibinfo{author}{\bibfnamefont{P.}~\bibnamefont{Ade}} \bibnamefont{et~al.}
  (\bibinfo{collaboration}{Simons Observatory})
  (\bibinfo{year}{2018}{\natexlab{b}}), \eprint{1808.07445}.

\bibitem[{\citenamefont{Ade et~al.}(2016)}]{Array:2015xqh}
\bibinfo{author}{\bibfnamefont{P.~A.~R.} \bibnamefont{Ade}}
  \bibnamefont{et~al.} (\bibinfo{collaboration}{BICEP2, Keck Array}),
  \bibinfo{journal}{Phys. Rev. Lett.} \textbf{\bibinfo{volume}{116}},
  \bibinfo{pages}{031302} (\bibinfo{year}{2016}), \eprint{1510.09217}.

\bibitem[{\citenamefont{Kofman et~al.}(1997)\citenamefont{Kofman, Linde, and
  Starobinsky}}]{Kofman:1997yn}
\bibinfo{author}{\bibfnamefont{L.}~\bibnamefont{Kofman}},
  \bibinfo{author}{\bibfnamefont{A.~D.} \bibnamefont{Linde}}, \bibnamefont{and}
  \bibinfo{author}{\bibfnamefont{A.~A.} \bibnamefont{Starobinsky}},
  \bibinfo{journal}{Phys. Rev.} \textbf{\bibinfo{volume}{D56}},
  \bibinfo{pages}{3258} (\bibinfo{year}{1997}), \eprint{hep-ph/9704452}.

\bibitem[{\citenamefont{Figueroa and Byrnes}(2017)}]{Figueroa:2016dsc}
\bibinfo{author}{\bibfnamefont{D.~G.} \bibnamefont{Figueroa}} \bibnamefont{and}
  \bibinfo{author}{\bibfnamefont{C.~T.} \bibnamefont{Byrnes}},
  \bibinfo{journal}{Phys. Lett.} \textbf{\bibinfo{volume}{B767}},
  \bibinfo{pages}{272} (\bibinfo{year}{2017}), \eprint{1604.03905}.

\bibitem[{\citenamefont{Nakama and Yokoyama}(2019)}]{Nakama:2018gll}
\bibinfo{author}{\bibfnamefont{T.}~\bibnamefont{Nakama}} \bibnamefont{and}
  \bibinfo{author}{\bibfnamefont{J.}~\bibnamefont{Yokoyama}},
  \bibinfo{journal}{PTEP} \textbf{\bibinfo{volume}{2019}},
  \bibinfo{pages}{033E02} (\bibinfo{year}{2019}), \eprint{1803.07111}.

\bibitem[{\citenamefont{Opferkuch et~al.}(2019)\citenamefont{Opferkuch,
  Schwaller, and Stefanek}}]{Opferkuch:2019zbd}
\bibinfo{author}{\bibfnamefont{T.}~\bibnamefont{Opferkuch}},
  \bibinfo{author}{\bibfnamefont{P.}~\bibnamefont{Schwaller}},
  \bibnamefont{and} \bibinfo{author}{\bibfnamefont{B.~A.}
  \bibnamefont{Stefanek}}, \bibinfo{journal}{JCAP}
  \textbf{\bibinfo{volume}{1907}}, \bibinfo{pages}{016} (\bibinfo{year}{2019}),
  \eprint{1905.06823}.

\bibitem[{\citenamefont{Dimopoulos and Markkanen}(2018)}]{Dimopoulos:2018wfg}
\bibinfo{author}{\bibfnamefont{K.}~\bibnamefont{Dimopoulos}} \bibnamefont{and}
  \bibinfo{author}{\bibfnamefont{T.}~\bibnamefont{Markkanen}},
  \bibinfo{journal}{JCAP} \textbf{\bibinfo{volume}{1806}}, \bibinfo{pages}{021}
  (\bibinfo{year}{2018}), \eprint{1803.07399}.

\bibitem[{\citenamefont{Haro}(2019)}]{Haro:2018jtb}
\bibinfo{author}{\bibfnamefont{J.}~\bibnamefont{Haro}}, \bibinfo{journal}{Phys.
  Rev.} \textbf{\bibinfo{volume}{D99}}, \bibinfo{pages}{043510}
  (\bibinfo{year}{2019}), \eprint{1807.07367}.

\bibitem[{\citenamefont{Hashiba and Yokoyama}(2019)}]{Hashiba:2018iff}
\bibinfo{author}{\bibfnamefont{S.}~\bibnamefont{Hashiba}} \bibnamefont{and}
  \bibinfo{author}{\bibfnamefont{J.}~\bibnamefont{Yokoyama}},
  \bibinfo{journal}{JCAP} \textbf{\bibinfo{volume}{1901}}, \bibinfo{pages}{028}
  (\bibinfo{year}{2019}), \eprint{1809.05410}.

\bibitem[{\citenamefont{Bettoni and Rubio}(2020)}]{Bettoni:2019dcw}
\bibinfo{author}{\bibfnamefont{D.}~\bibnamefont{Bettoni}} \bibnamefont{and}
  \bibinfo{author}{\bibfnamefont{J.}~\bibnamefont{Rubio}},
  \bibinfo{journal}{JCAP} \textbf{\bibinfo{volume}{2001}}, \bibinfo{pages}{002}
  (\bibinfo{year}{2020}), \eprint{1911.03484}.

\bibitem[{\citenamefont{Watanabe and Komatsu}(2007)}]{Watanabe:2006ku}
\bibinfo{author}{\bibfnamefont{Y.}~\bibnamefont{Watanabe}} \bibnamefont{and}
  \bibinfo{author}{\bibfnamefont{E.}~\bibnamefont{Komatsu}},
  \bibinfo{journal}{Phys. Rev.} \textbf{\bibinfo{volume}{D75}},
  \bibinfo{pages}{061301} (\bibinfo{year}{2007}), \eprint{gr-qc/0612120}.

\bibitem[{\citenamefont{Chun et~al.}(2009)\citenamefont{Chun, Scopel, and
  Zaballa}}]{Chun:2009yu}
\bibinfo{author}{\bibfnamefont{E.~J.} \bibnamefont{Chun}},
  \bibinfo{author}{\bibfnamefont{S.}~\bibnamefont{Scopel}}, \bibnamefont{and}
  \bibinfo{author}{\bibfnamefont{I.}~\bibnamefont{Zaballa}},
  \bibinfo{journal}{JCAP} \textbf{\bibinfo{volume}{0907}}, \bibinfo{pages}{022}
  (\bibinfo{year}{2009}), \eprint{0904.0675}.

\bibitem[{\citenamefont{Hasegawa et~al.}(2019)\citenamefont{Hasegawa,
  Hiroshima, Kohri, Hansen, Tram, and Hannestad}}]{Hasegawa:2019jsa}
\bibinfo{author}{\bibfnamefont{T.}~\bibnamefont{Hasegawa}},
  \bibinfo{author}{\bibfnamefont{N.}~\bibnamefont{Hiroshima}},
  \bibinfo{author}{\bibfnamefont{K.}~\bibnamefont{Kohri}},
  \bibinfo{author}{\bibfnamefont{R.~S.} \bibnamefont{Hansen}},
  \bibinfo{author}{\bibfnamefont{T.}~\bibnamefont{Tram}}, \bibnamefont{and}
  \bibinfo{author}{\bibfnamefont{S.}~\bibnamefont{Hannestad}},
  \bibinfo{journal}{JCAP} \textbf{\bibinfo{volume}{12}}, \bibinfo{pages}{012}
  (\bibinfo{year}{2019}), \eprint{1908.10189}.

\bibitem[{\citenamefont{Carroll}(1998)}]{Carroll:1998zi}
\bibinfo{author}{\bibfnamefont{S.~M.} \bibnamefont{Carroll}},
  \bibinfo{journal}{Phys. Rev. Lett.} \textbf{\bibinfo{volume}{81}},
  \bibinfo{pages}{3067} (\bibinfo{year}{1998}), \eprint{astro-ph/9806099}.

\bibitem[{\citenamefont{Dvali and Zaldarriaga}(2002)}]{Dvali:2001dd}
\bibinfo{author}{\bibfnamefont{G.~R.} \bibnamefont{Dvali}} \bibnamefont{and}
  \bibinfo{author}{\bibfnamefont{M.}~\bibnamefont{Zaldarriaga}},
  \bibinfo{journal}{Phys. Rev. Lett.} \textbf{\bibinfo{volume}{88}},
  \bibinfo{pages}{091303} (\bibinfo{year}{2002}), \eprint{hep-ph/0108217}.

\bibitem[{\citenamefont{Chiba and Kohri}(2002)}]{Chiba:2001er}
\bibinfo{author}{\bibfnamefont{T.}~\bibnamefont{Chiba}} \bibnamefont{and}
  \bibinfo{author}{\bibfnamefont{K.}~\bibnamefont{Kohri}},
  \bibinfo{journal}{Prog. Theor. Phys.} \textbf{\bibinfo{volume}{107}},
  \bibinfo{pages}{631} (\bibinfo{year}{2002}), \eprint{hep-ph/0111086}.

\end{thebibliography}

\end{document}